\documentclass[12pt, prd]{revtex4}

\usepackage{amsmath}
\usepackage{amssymb}
\usepackage[dvips]{graphicx,color}

\newcommand{\Slash}[1]{{\ooalign{\hfil#1\hfil\crcr\raise.167ex\hbox{/}}}}

\begin{document}

\title{
Neutral bions in the ${\mathbb C}P^{N-1}$ model 
}

\author{Tatsuhiro Misumi}
\email{misumi(at)phys-h.keio.ac.jp}
\affiliation{Department of Physics, and Research and Education 
Center for Natural Sciences, 
Keio University, Hiyoshi 4-1-1, Yokohama, Kanagawa 223-8521, Japan}

\author{Muneto Nitta}
\email{nitta(at)phys-h.keio.ac.jp}
\affiliation{Department of Physics, and Research and Education Center for Natural Sciences, 
Keio University, Hiyoshi 4-1-1, Yokohama, Kanagawa 223-8521, Japan}

\author{Norisuke Sakai}
\email{norisuke.sakai(at)gmail.com}
\affiliation{Department of Physics, and Research and Education Center for Natural Sciences, 
Keio University, Hiyoshi 4-1-1, Yokohama, Kanagawa 223-8521, Japan}

\begin{abstract}
We study classical configurations in the ${\mathbb C}P^{N-1}$ model 
on ${\mathbb R}^{1}\times S^{1}$ with twisted boundary conditions.
We focus on specific configurations composed of multiple 
fractionalized-instantons, termed ``neutral bions", which are 
identified as ``perturbative infrared renormalons" 
by \"{U}nsal and his collaborators.
For ${\mathbb Z}_N$ twisted boundary conditions, 
we consider an explicit ansatz corresponding to topologically 
trivial configurations 
containing one fractionalized instanton ($\nu=1/N$) 
and one fractionalized anti-instanton ($\nu=-1/N$) at large separations,
and exhibit the attractive interaction between the instanton constituents 
and how they behave at shorter separations.
We show that the bosonic interaction potential between the constituents as 
a function of both the separation and $N$ 
is consistent with the standard separated-instanton calculus 
even from short to large separations, 
which indicates that the ansatz enables us to study bions 
and the related physics for a wide range of separations.
We also propose different bion ansatze in a certain 
non-${\mathbb Z}_{N}$ twisted boundary condition corresponding 
to the ``split" vacuum for $N= 3$ 
and its extensions for $N\geq 3$.
We find that the interaction potential
has qualitatively the same asymptotic behavior and $N$-dependence 
as those of bions for ${\mathbb Z}_{N}$ twisted boundary conditions.
\end{abstract}

\maketitle

\newpage


\section{Introduction}
\label{sec:Intro}

In the recent study on QCD-like theories with spatial compactification ($L$), 
fractionalized multi-instanton configurations composed of 
fractionalized instantons and anti-instantons have been attracting a great deal of attention. 
It is stressed by \"{U}nsal and his collaborators that these configurations 
\cite{Yu1, RS1, U1, U2, SU1, PU2, PSU1, AU1, DU1, DD1, DU2, CDDU1, BDU1,DU3,CDU1}, 
which are termed ``bions", have two physical significances associated with 
two types of topologically trivial bion configurations called ``magnetic 
(charged) bions" and ``neutral bions", as seen in the following examples: 
In the weak-coupling regime ($L\ll1/\Lambda_{\rm QCD}$) in QCD(adj.) 
on ${\mathbb R}^{3} \times S^{1}$, or in the $U(1)^{N-1}$ center-symmetric 
phase \cite{H1, H2, MO1, MO3, CD1, MeO1, NO1, O1, KM1, CHHN1}, 
condensation of magnetic bions (zero topological charge and nonzero magnetic 
charge) causes the confinement \cite{U1, U2, SU1, PU2, PSU1}. 
This confinement mechanism may remain responsible for the confinement
at strong-coupling regime due to the continuity principle.
This argument is also of importance in terms of the recent progress in
large-$N$ volume reduction \cite{EK1, KUY1, UY1, BS1, B1, BS2, B2, PU1, AHUY1,BKS1}.
On the other side, neutral bions (zero topological charge 
and zero magnetic charge) can be identified as 
the infrared renormalon \cite{AU1, DU1, DD1, DU2, CDDU1, BDU1, DU3, CDU1,tH1, FKW1}.
Here imaginary ambiguities arising in bion's amplitude and those arising in non-Borel-summable perturbative series cancel against each other, and it is expected that full semi-classical expansion including perturbative and non-perturbative sectors,
which is called ``resurgent" expansion \cite{Ec1}, leads to unambiguous and self-consistent definition of field theories in the same manner as the Bogomolny-Zinn-Justin (BZJ) 
prescription in quantum mechanics \cite{Bogomolny:1980ur, 
ZinnJustin:1981dx, ZinnJustin:2004ib}. 
However, it is not straightforward to verify these arguments in gauge 
theories directly, since it is difficult to find an explicit ansatz of bion configurations.

In order to reach deeper understanding on bions and 
the associated physics, it is of great importance to study
examples in the low-dimensional models such as 
${\mathbb C}P^{N-1}$ models \cite{DU1, DD1}, 
principal chiral models \cite{CDDU1,CDU1} and quantum mechanics \cite{DU2, BDU1, DU3}. 
In particular, the ${\mathbb C}P^{N-1}$ model in 1+1 dimensions has 
been studied for a long time as a toy model 
of the Yang-Mills theory in 3+1 dimensions \cite{Polyakov}, 
because of similarities between them such as 
dynamical mass gap, 
asymptotic freedom and the existence of instantons \cite{Polyakov:1975yp}.
The ${\mathbb C}P^{N-1}$ model  
on ${\mathbb R}^1 \times S^1$ with twisted boundary conditions
admits fractionalized instantons (domain wall-instantons) 
as configurations with the minimal topological charge 
\cite{Eto:2004rz,Eto:2006mz} 
(see also Refs.~\cite{Bruckmann:2007zh}). 
In Ref.~\cite{DU1}, generic arguments on bion configurations were given 
in the ${\mathbb C}P^{N-1}$ model on ${\mathbb R}^1 \times S^1$
with ${\mathbb Z}_{N}$ twisted boundary conditions, which is a 
corresponding situation to $U(1)^{N-1}$ center-symmetric phase in QCD(adj.),
based on the independent instanton description taking account of 
interactions between far-separated fractionalized 
instantons and anti-instantons.
According to the study, the renormalon ambiguity 
arising in non-Borel-summable perturbative series is compensated by the amplitude of 
neutral bions also in the ${\mathbb C}P^{N-1}$ model. 
This phenomenon, which is called ``resurgence", works as follows \cite{DU1}:
The effective interaction action by bosonic exchange between one fractionalized instanton $\mathcal{K}_{i}$ and one fractionalized anti-instanton $\overline{\mathcal{K}}_{j}$ is
\begin{equation}
S_{\rm int}(\tau)=-4\xi {\alpha_{i}\cdot \alpha_{j}\over{g^{2}}}e^{-\xi\tau}\,,
\,\,\,\,\,\,\,\,\,\,\,\,\,
\xi\equiv {2\pi\over{N}}\,,
\label{bionS}
\end{equation}
where $\tau$ stands for distance (divided by the compact scale $L$) 
between two fractionalized instantons. 
Vectors $\alpha_{i}, \alpha_{j}$ are affine co-roots 
and $\alpha_{i}\cdot\alpha_{j}$ is an entry of the extended Cartan matrix.
The total bion amplitude including the fermion zero-mode exchange contribution 
is mainly given by
\begin{equation}
\mathcal{B}_{ij} \propto -e^{-2S_{I}/N}\int_{0}^{\infty} d\tau\, e^{-V_{\rm eff}^{ij}(\tau)}\,,
\end{equation}
with $V_{\rm eff}^{ij}(\tau)= S_{\rm int}(\tau)+2N_{f}\xi\tau$ and $S_{I}$ being 
the instanton action. $N_{f}$ stands for fermion flavors.
For neutral bion $\alpha_{i}\cdot\alpha_{i}>0$, 
semiclassical description of independent fractionalized instantons breaks down 
since the interaction is attractive and instantons are merged 
in the end. 
Here, the BZJ-prescription, replacing $g^{2}\to-g^{2}$, 
works to extract meaningful information from this amplitude.
The prescription turns the interaction (spuriously) into a 
repulsive one and the amplitude becomes well-defined as
\begin{equation}
\mathcal{B}_{ii}(g^{2}, N_{f})\,\,\,\to\,\,\,\tilde{\mathcal{B}}_{ii}
(-g^{2}, N_{f})\,\,\,\propto\,\,\, 
(-g^{2}N/8\pi)^{2N_{f}}\Gamma(2N_{f})e^{-2S_{I}/N}\,.
\end{equation}
By the use of the analytic continuation in 
the $g^{2}$ complex plane, we can continue back to the original $g^{2}$.
For $N_{f}=0$ case, we then encounter the following imaginary ambiguity in the amplitude as
\begin{equation}
\tilde{\mathcal{B}}_{ii}(g^{2}, 0) \,\,\,\propto\,\,\,\left( \log(g^{2}N/8\pi)-\gamma \pm i\pi\right)
e^{-2S_{I}/N}\,.
\end{equation}
We can rephrase this situation as unstable negative modes of bions give 
rise to imaginary ambiguities of the amplitude.
The imaginary ambiguity has the same magnitude with an opposite sign 
as the leading-order ambiguity ($\sim \mp i\pi e^{-2S_{I}/N}$) arising from 
the non-Borel-summable series expanded around the perturbative vacuum.
The ambiguities at higher orders ($\mp i\pi e^{-4S_{I}/N}$, $\mp i\pi e^{-6S_{I}/N}$,...) 
are cancelled by amplitudes of bion molecules (2-bion, 3-bion,...), 
and the full trans-series expansion around 
the perturbative and non-perturbative vacua results in unambiguous 
definition of field theories.

Although this generic argument based on far-separated instantons 
is clear, it is also worthwhile manifesting and studying an 
explicit solution or ansatz corresponding to bion configurations, 
which can be investigated from short to large separation. 
In Ref.~\cite{DD1}, the authors found out 
non-Bogomol'nyi-Prasad-Sommerfield (BPS) solutions 
in the ${\mathbb C}P^{N-1}$ model on ${\mathbb R}^1 \times S^1$
with a ${\mathbb Z}_{N}$ twisted boundary condition, 
and have shown that these solutions can be critical points, 
around which the resurgent semi-classical expansion is performed.
The simplest non-BPS solution that they found is a four-instanton 
configuration composed of two fractionalized instantons ($\nu=1/N$) 
and two fractionalized anti-instantons ($\nu=-1/N$) for $N\geq3$.
(see also \cite{Bolognesi:2013tya}.)
They used the Din-Zakrewski projection method \cite{Zak1} 
generating a tower of non-BPS solutions from a BPS solution. 
It is known that all possible classical solutions are exhausted 
by this method at least on ${\mathbb R}^{2}$ and $S^2$ \cite{Zak1}.
This result indicates that if a simple bion configuration 
containing one instanton and one anti-instanton 
in the ${\mathbb C}P^{N-1}$ model ($N\geq2$) exists, 
it may not be a solution of the equation of motion,  
but may be some classical configuration which can give 
significant contributions to path integrals.
If it is true, one question arises what ansatz corresponds to such a bion. 
If such an ansatz exists, the other questions arise 
how the instanton constituents behave at short separations and
whether it is consistent with the amplitude (\ref{bionS}) obtained in 
the standard instanton calculus in a far-separated limit.
In the present study, we consider and study an ansatz corresponding 
to bions beyond exact solutions.
We also consider more general twisted boundary conditions 
similar to the ``split phase" in QCD(adj.).

The purpose of our work is to study an explicit ansatz corresponding to topologically trivial bion configurations in the ${\mathbb C}P^{N-1}$
on ${\mathbb R}^{1}\times S^{1}$ with several twisted boundary conditions,
and show how the instanton constituents behave at an arbitrary separation.
For the ${\mathbb Z}_{N}$ twisted boundary condition, 
we consider a simple neutral-bion ansatz, which yields configuration 
involving one fractionalized instanton ($\nu=1/N$) and one 
fractionalized anti-instanton ($\nu=-1/N$) in 
the well-separated limit.
By studying separation dependence of the total action, 
we exhibit the attractive interaction between the instanton 
constituents and how they are merged in the end, which means 
that the configuration has a negative mode.
By looking into $N$-dependence of the interaction potential 
as a function of the separation in comparison with the result 
in the standard instanton calculus (\ref{bionS}), 
we show that our ansatz is consistent with (\ref{bionS}) even 
from short to large separations. 
Our ansatz can be used to study bions and related physics 
for a wide range of separations.
For the non-${\mathbb Z}_{N}$ twisted boundary conditions with $N=3$, 
which we term a ``split" boundary condition, 
we find out a different fractionalized instanton-anti-instanton ansatz.
We again show that the configuration has a negative mode.
We extend the ansatz to general $N\geq 3$ cases, and find 
that the interaction potential between the instantons has 
qualitatively the same properties as (\ref{bionS}) up to 
some factors in the extended versions. 
This fact indicates universality of resurgence based on neutral bions 
for general boundary conditions.

In Sec.~\ref{sec:CP} we introduce ${\mathbb C}P^{N-1}$ models 
with some notations for calculations.
In Sec.~\ref{sec:ZN} we first introduce ${\mathbb Z}_{N}$ 
twisted boundary conditions and discuss how fractionalized instantons emerge.
We then propose a specific ansatz for neutral bions, and 
discuss the properties.
In Sec.~\ref{sec:oTBC} we consider non-${\mathbb Z}_{N}$ twisted 
boundary conditions, and discuss bion-like configurations for the cases.
Section \ref{sec:SD} is devoted to a summary.


\section{${\mathbb C}P^{N-1}$ model}
\label{sec:CP}

Let $\omega(x)
$ be an $N$-component vector of 
complex scalar fields, and $n(x)
$ be 
a normalized complex $N$-component vector 
composed from $\omega$: 
$n(x) \equiv \omega(x)/|\omega(x)|$ with 
$|\omega|=\sqrt{\omega^\dagger \omega}$. 
Then, the action and topological charge 
 representing $\pi_2({\mathbb C}P^{N-1}) \simeq {\mathbb Z}$ 
of the ${\mathbb C}P^{N-1}$ model 
in Euclidean two dimensions 
are given by (see, e.g., Ref.~\cite{Zak1})
\begin{align}
S&={1\over{g^{2}}}\int d^{2}x (D_{\mu}n)^{\dag} (D_{\mu} n)\,,
\\
Q&=\int d^{2}x \; i\epsilon_{\mu\nu} (D_{\nu}n)^{\dag} (D_{\mu} n)=\int d^{2}x \epsilon_{\mu\nu}\partial_{\mu} A_{\nu}\,,
\label{Qdef}
\end{align}
respectively, 
where $d^{2} x\equiv dx_{1}dx_{2}$ and $\mu,\nu=1,2$.
Here, we have defined the covariant derivative by  $D_{\mu}=\partial_{\mu}-iA_{\mu}$ 
with a composite gauge field $A_{\mu} (x) \equiv -in^{\dag}\partial_{\mu}n$.

The action $S$ and topological charge $Q$ can be expressed 
in terms of the projection operator 
${\bf P}\equiv nn^{\dag}= {\omega \omega^{\dag}\over{\omega^{\dag}\omega}}$ 
and using the complex coordinate $z \equiv x_1+ix_2$, 
\begin{align}
S&={2\over{g^{2}}}\int d^2 x {\rm Tr} \left[ \partial_{z}{\bf P}\partial_{\bar z} {\bf P} \right]\,,
\\
Q&=2\int d^2 x {\rm Tr} \left[{\bf P} (\partial_{\bar z}{\bf P}\partial_{z} {\bf P}-\partial_{z}{\bf P}\partial_{\bar z} {\bf P}) \right]\,.
\end{align}

All through this paper, we focus the geometry ${\mathbb R}^{1}\times S^{1}$ 
and configurations on it satisfying periodicity in the $x_2$ 
direction with compactification scale $L$.
For all the configurations considered in the present paper, 
the action density and the topological charge density 
are reduced to be functions of $x_{1}$:
\begin{align}
S&=\int dx_{1}\, s(x_{1})
={1\over{g^{2}\pi}}\int d^{2}x \,{\rm Tr} 
\left[ \partial_{z}{\bf P}\partial_{\bar z} {\bf P} \right]\,,
\\
Q&=\int dx_{1} \,q(x_{1})
={1\over{\pi}}\int d^{2}x\, 
{\rm Tr} \left[{\bf P} 
(\partial_{\bar z}{\bf P}\partial_{z} {\bf P}
-\partial_{z}{\bf P}\partial_{\bar z} {\bf P}) \right]\,,
\end{align}
where we have defined the action density $s(x_{1})$ and the charge 
density $q(x_{1})$ depending only on $x_{1}$. 
Here we redefine the action and topological charge as $S/(2\pi) \to S$ and $Q/(2\pi) \to Q$
for them to have multiples of $1/N$ after $x_{1}$ integration. 
In this paper, we omit the coupling $1/g^2$ for simplicity.

The ${\mathbb C}P^1$ model is equivalent to 
the $O(3)$ nonlinear sigma model, 
described by three real scalar fields 
${\bf m}(x)=(m^1(x),m^2(x),m^3(x))^T$ 
with a constraint ${\bf m}(x)^2=1$.
More explicitly, 
\begin{eqnarray}
  {\bf m}(x)&=& 
 n^\dagger(x) \vec{\sigma} n(x) 
=
\frac{\omega^\dagger(x) \vec{\sigma} \omega(x)}
{\omega^\dagger(x) \omega(x)} 
\\ \nonumber
&=& \frac{(\omega^{*1}\omega^2 + \omega^{*2}\omega^1,
-i\omega^{*1}\omega^2 + i \omega^{*2}\omega^1,
|\omega^1|^2 - |\omega^2|^2)}
{\omega^\dagger(x) \omega(x)} 
,
\label{eq:three-vector}
\end{eqnarray}
with the Pauli matrices $\vec{\sigma}$. 
Then, the action is
\begin{align}
 S = \frac{1}{g^2}
\int d^2x \partial_{\mu} {\bf m} \cdot  \partial_{\mu} {\bf m} .
\end{align}


\section{Fractionalized instantons and neutral-bion configuration in ${\mathbb Z}_{N}$ twisted boundary conditions}
\label{sec:ZN}

\subsection{${\mathbb Z}_{N}$ twisted boundary conditions}
In the present section, we propose a neutral bion ansatz for a 
${\mathbb Z}_{N}$ twisted boundary condition 
in the ${\mathbb C}P^{N-1}$ model  
on ${\mathbb R}^1 \times S^1$.
${\mathbb Z}_{N}$ twisted boundary conditions in a compactified direction is expressed as
\cite{DU1, DD1}
\begin{equation}
\omega(x_{1}, x_{2}+L) = \Omega \,\omega(x_{1},x_{2})\,,
\,\,\,\,\,\,\,\,\,\,\,\,\,
\Omega={\rm diag.}\left[1, e^{2\pi i/N}, e^{4\pi i/N},\cdot\cdot\cdot, e^{2(N-1)\pi i/N}  \right]\,.
\label{ZNC}
\end{equation}
In $SU(N)$ gauge theories with adjoint quarks, this 
${\mathbb Z}_{N}$ twisted boundary condition 
corresponds to the vacuum with the gauge symmetry breaking 
$SU(N) \to U(1)^{N-1}$, 
where Wilson-loop holonomy in the compactified direction is given by
\begin{equation}
\langle A_{2} \rangle = (0, 2\pi/N, \cdot\cdot\cdot, 
2(N-1)\pi/N)\,,\,\,\,\,\,\,\,\,\,{\rm for}\,\,\,\,\,N\geq3\,,
\label{WHN}
\end{equation}
and
\begin{equation}
\langle A_{2} \rangle = (-\pi/2, \pi/2)\,,\,\,\,\,\,\,\,
\,\,\,\,\,\,\,\,\,\,{\rm for}\,\,\,\,\,N=2\,,
\label{WHN2}
\end{equation}
where $A_{2}$ is the gauge field in the compactified direction.
(See also \cite{KSMTSY1, SKSY1, KMSSY1, KMKMSY1}
for topics related to ${\mathbb Z}_{N}$ twisted boundary conditions.)
We here omit permutation copies. 
We note that the gauge field defined in the ${\mathbb C}P^{N-1}$ 
model (\ref{Qdef}) also has 
the same Wilson-loop holonomy for ${\mathbb Z}_{N}$ twisted boundary condition.
Difference between (\ref{ZNC}) and (\ref{WHN2}) for $N=2$ is just 
superficial and unphysical,
since two different ansatz of $\omega(x)$ with an overall boundary 
condition factor $e^{-i \pi/2}$ result in the same projection 
field ${\bf P}(x)$ as we will show later.
Fractionalized instantons (domain wall-instantons) 
carry the minimum topological charges
in the ${\mathbb C}P^{N-1}$ model  
on ${\mathbb R}^1 \times S^1$ with a 
twisted boundary condition 
\cite{Eto:2004rz,Eto:2006mz}.
For simplicity, we begin with the ${\mathbb C}P^{1}$ model and generalize the argument 
to the ${\mathbb C}P^{N-1}$ model subsequently.
From next subsection we make all the dimensionful quantities and parameters dimensionless by using the compact scale $L$ ($L\to 1$) unless we have a special reason to recover it.

\subsection{Fractionalized instantons}

In this subsection, we illustrate fractionalized instantons 
in the ${\mathbb C}P^{1}$ model 
 satisfying a ${\mathbb Z}_{2}$ twisted boundary condition (\ref{ZNC}) as 
\begin{eqnarray}
&& \omega (x_1,x_2+1
) = {\rm diag.}[1,e^{\pi i}] \omega (x_1,x_2)
= {\rm diag.}[1,-1] \omega (x_1,x_2),  \\
&& (m^1(x_1,x_2+1),m^2(x_1,x_2+1),m^3(x_1,x_2+1)) 
\nonumber \\ 
&& = (-m^1(x_1,x_2),-m^2(x_1,x_2),m^3(x_1,x_2+1))\,, 
\end{eqnarray}
on ${\mathbb R}^{1}\times S^{1}$ with 
the unexplicit unit compactification scale $L$ \cite{footnote}.
Here, we have used the relation (\ref{eq:three-vector}) 
for the second equation.

\begin{figure}[htbp]
\begin{center}
\begin{tabular}{cc}
 \includegraphics[width=0.3\textwidth]{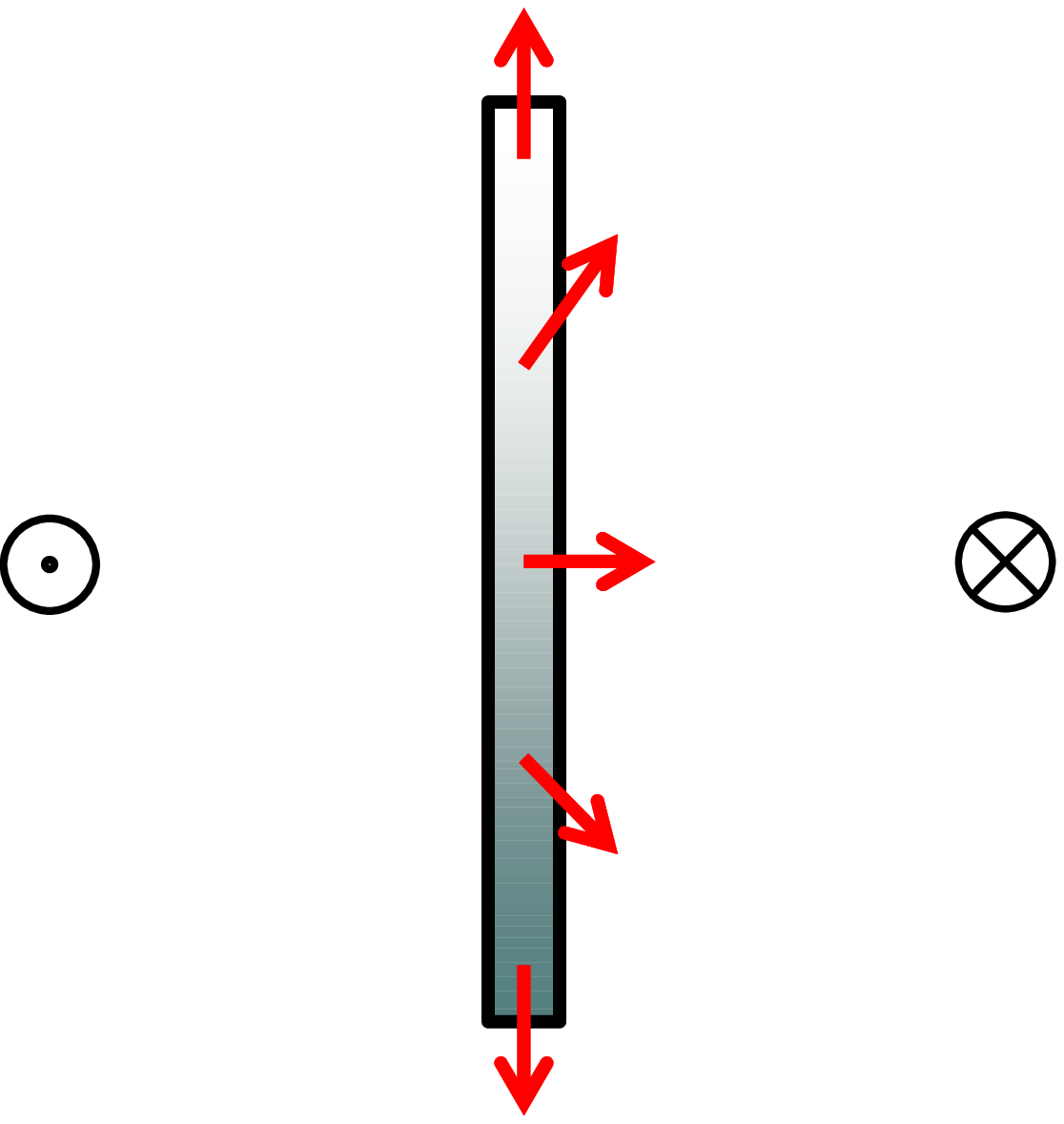}
& \includegraphics[width=0.3\textwidth]{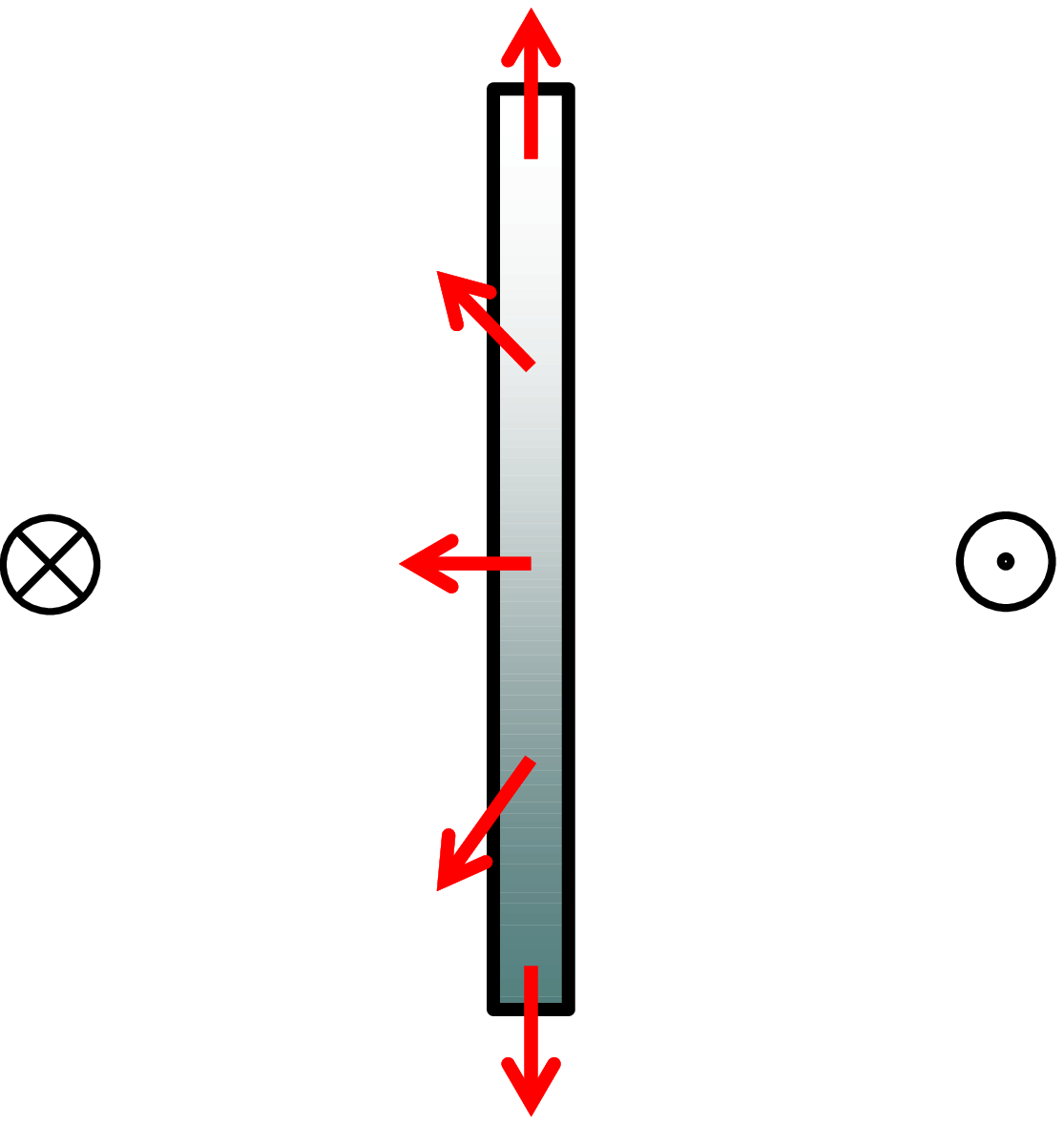}\\
(a) $\omega_L$ & (b) $\omega_R$\\
 \includegraphics[width=0.3\textwidth]{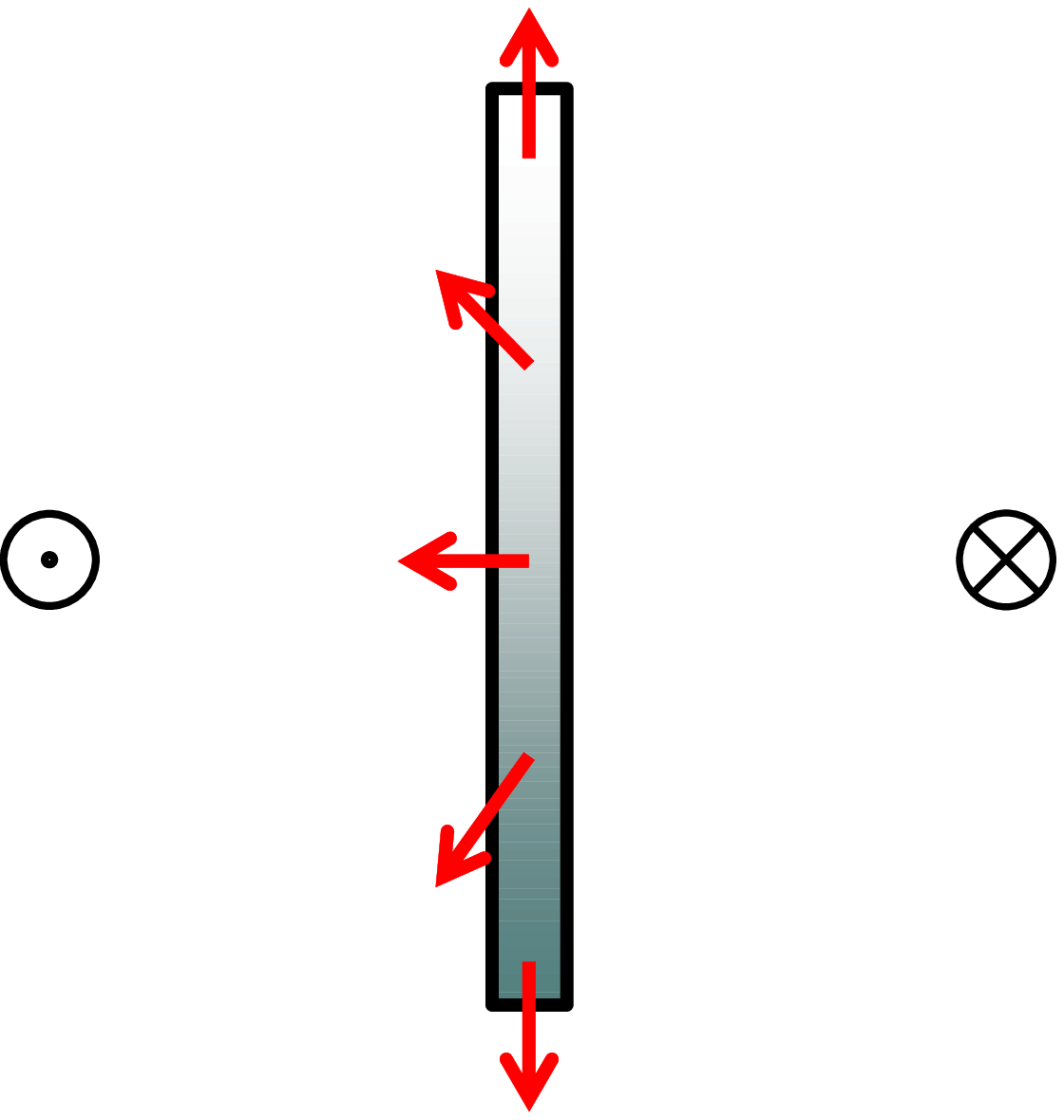}
& \includegraphics[width=0.3\textwidth]{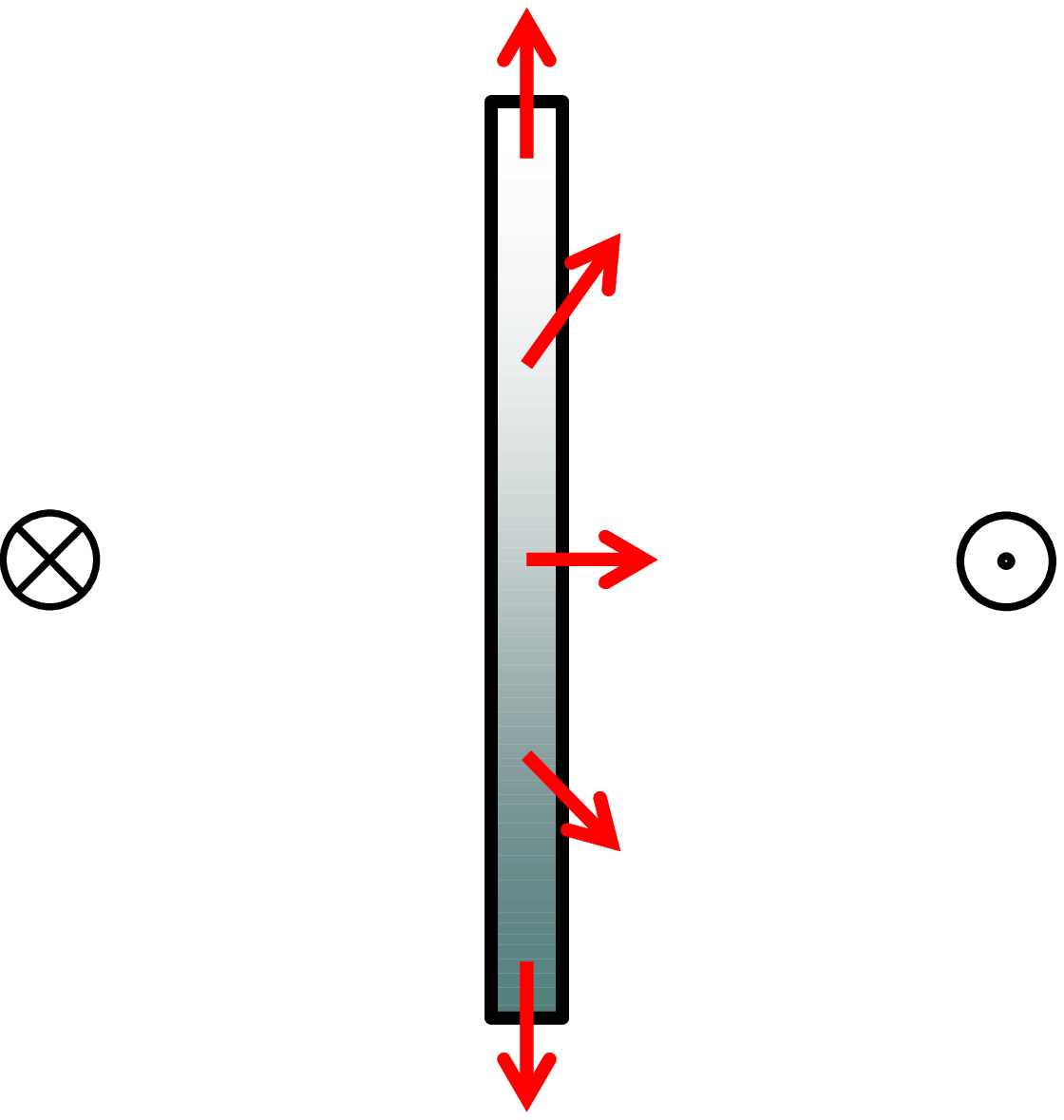}\\
(c) $\omega_L^*$ & (d) $\omega_R^*$
\end{tabular}
\end{center}
\caption{
Fractionalized instantons in 
the ${\mathbb C}P^1$ model with the ${\mathbb Z}_2$ twisted 
boundary condition, 
corresponding to (a) $\omega_L$, (b) $\omega_R$, 
(c) $\omega_L^*$, and (d) $\omega_R^*$ 
(in which we have taken the phase modulus to be 
$\theta = - \pi/2$). 
The horizontal and vertical directions are 
$x_1$ and $x_2$, respectively.
The symbols 
$\odot$, $\otimes$, $\leftarrow$, $\rightarrow$, 
$\uparrow$ and $\downarrow$ 
denote 
${\bf m}=(0,0,1),(0,0,-1),(-1,0,0),(1,0,0),(0,1,0)$ and 
$(0,-1,0)$, respectively. 
The shaded regions imply domain walls 
with $m^3 \sim 0$. 
The $\uparrow$ and $\downarrow$ at the boundaries 
at $x_2=+1$ and $x_2=0$ are identified 
by the twisted boundary condition. 
The domain wall charges are (a) $+1$, (b) $-1$, (c) $+1$, (d) $-1$,
and the instanton charges $Q$ are  (a) $+1/2$, (b) $+1/2$, 
(c) $-1/2$, (d) $-1/2$.
}
\label{fig:fractional-config}
\end{figure}

\begin{figure}[htbp]
\begin{center}
\begin{tabular}{ccc}
 \includegraphics[width=0.33\textwidth]{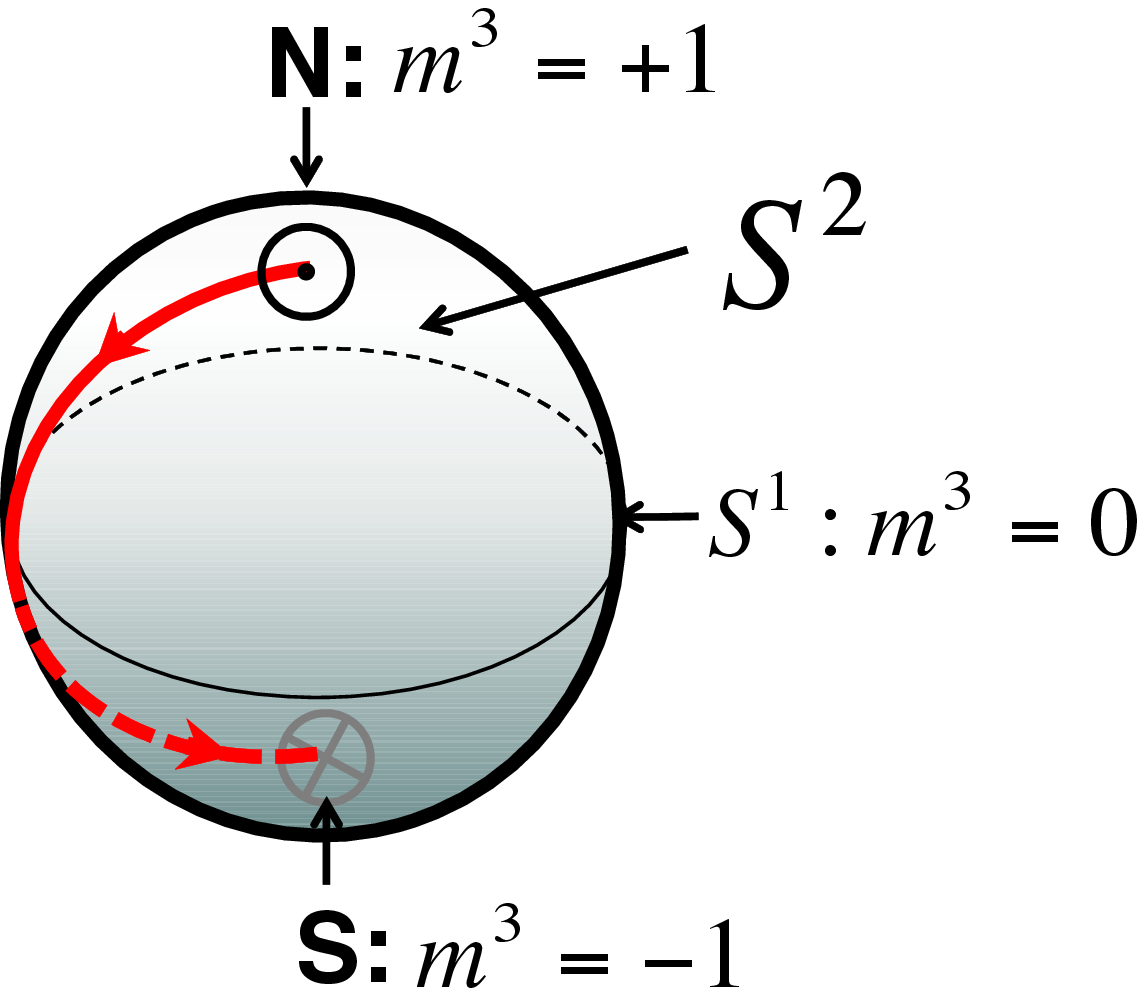} &
 \includegraphics[width=0.33\textwidth]{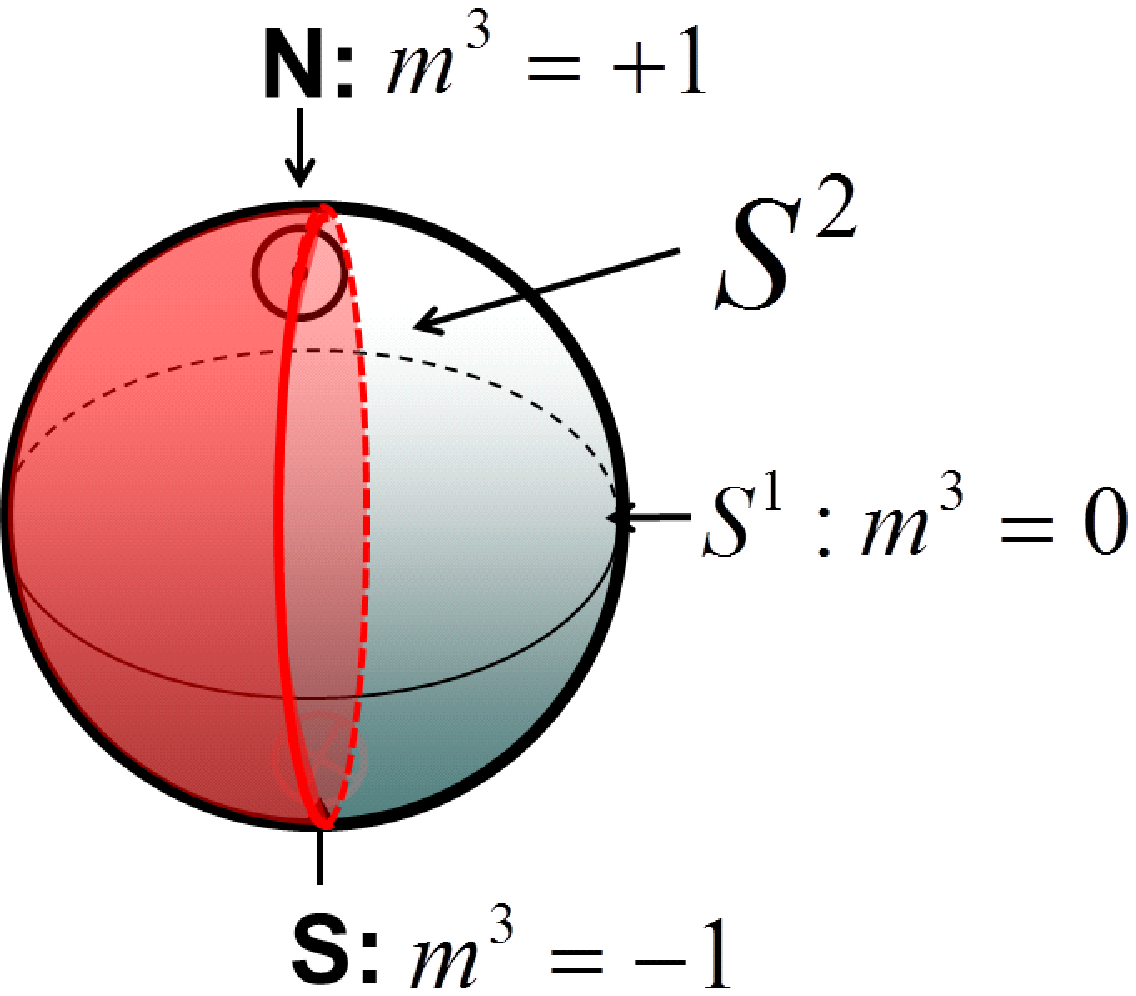}
& \includegraphics[width=0.33\textwidth]{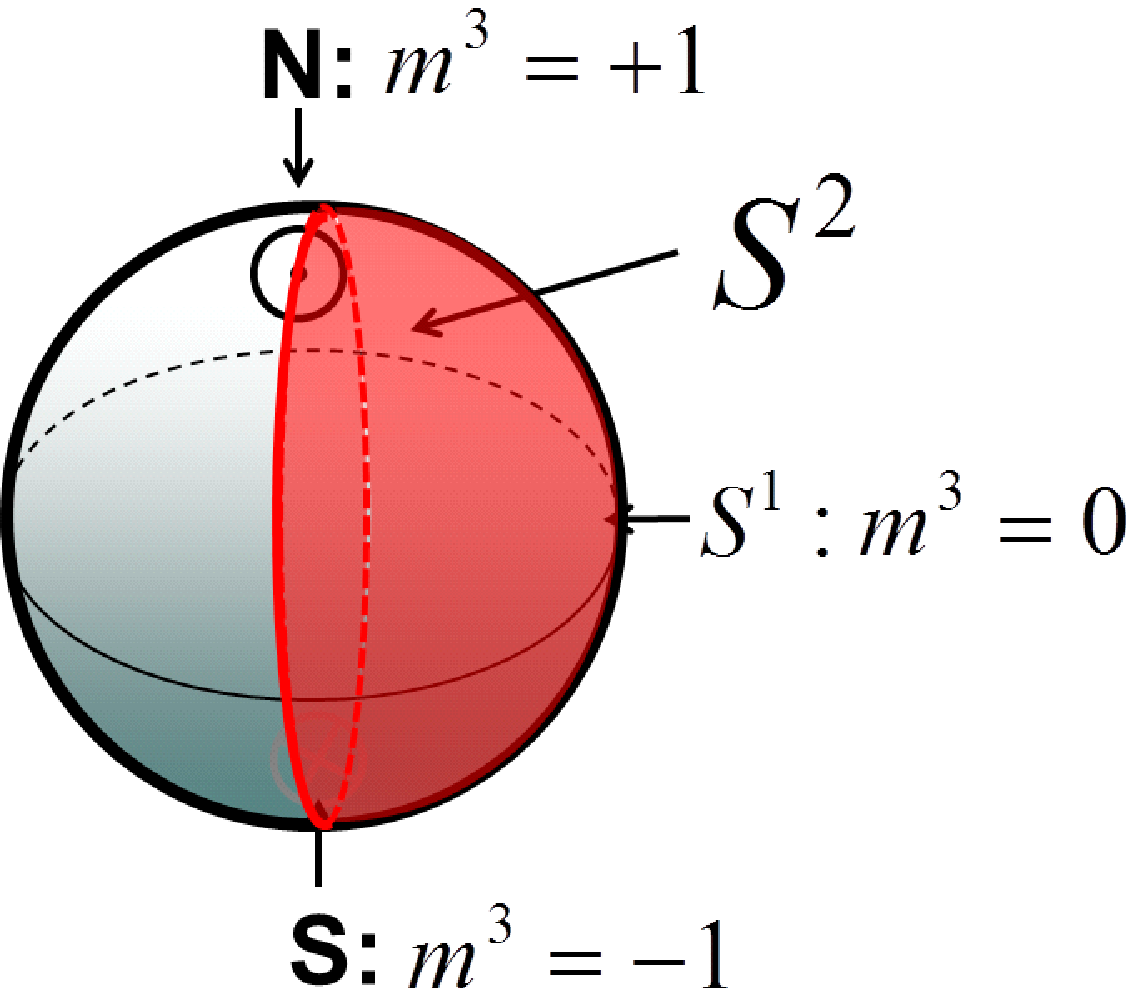}\\
(a) & (b) & (c)
\end{tabular}
\end{center}
\caption{
(a) Domain wall, and 
(b) and (c) fractionalized instantons 
in the target space $S^2$.
(b) corresponds to
the configurations $\omega_R$ and $\omega_L^*$ 
while (c) corresponds to 
the configurations $\omega_L$ and $\omega_R^*$.
}
\label{fig:fractional-sphere}
\end{figure}

Using the complex coordinate $z = x_{1}+ix_{2}$ on
${\mathbb R}^{1}\times S^{1}$,
fractionalized instanton solutions are given by 
\begin{eqnarray}
&& \omega_L
 = \left(1,
\lambda e^{i\theta}e^{+\pi z} \right)^{T}\,, \quad
 \omega_R
 = \left(1,
\lambda e^{i\theta}e^{-\pi z} \right)^{T}\,, \nonumber \\
&& \omega_L^*
 = \left(1,
\lambda e^{i\theta}e^{+\pi \bar z} \right)^{T}\,, \quad
 \omega_R^*
 = \left(1,
\lambda e^{i\theta}e^{-\pi \bar z} \right)^{T}\,, 
 \label{eq:fractional}
\end{eqnarray}
with real constants $\lambda$ and $\theta$ 
which are moduli. 
The configurations $\omega_L$  and $\omega_R$ are BPS 
which are holomorphic and depend on $z$,  
while their complex conjugate 
$\omega_L^*$  and $\omega_R^*$ are anti-BPS 
which are anti-holomorphic and depend on $\bar z$ only.
Fig.~\ref{fig:fractional-config} shows 
configurations in ${\bf m}(x)$ of these solutions.
The configuration $\omega_L$ ($\omega_L^*$)
goes to $n = (1,0)$ 
(${\bf m}=(0,0,1)$) denoted by $\odot$ at $x_1 \to -\infty$
and to 
$n = (0,1)$ 
(${\bf m}=(0,0,-1)$) denoted by $\otimes$ at $x_1 \to +\infty$.
The configuration $\omega_R$ ($\omega_R^*$)
goes to $n = (0,1)$ 
(${\bf m}=(0,0,-1)$) at $x_1 \to -\infty$
and to 
$\omega = (1,0)$ 
(${\bf m}=(0,0,+1)$) at $x_1 \to +\infty$.
The configurations $\omega_L$ ($\omega_L^*$) and $\omega_R$ 
($\omega_R^*$) can be regarded  as 
a domain wall and anti-domain wall, respectively.   
A domain wall 
at each constant $x_2$ slice 
corresponds to a path connecting 
the north pole ${\bf m}=(0,0,+1)$ and the south pole 
 ${\bf m}=(0,0,-1)$ in the target space, 
as illustrated in Fig.~\ref{fig:fractional-sphere}(a).  
A $U(1)$ modulus is localized 
on these domain walls   
characterizing which point on the equator 
in the target space 
a domain wall passes through \cite{Arai:2002xa}.
This $U(1)$ modulus is twisted along the 
domain wall to satisfy the boundary condition 
at $x_2 = 0$ and $x_2 =1$. 
When one changes a constant $x_2$ slice  
from $x_2=0$ to $x_2=1$, 
a path in the target space changes 
with sweeping a half of the sphere as the target space,  
as shown in Fig.\ref{fig:fractional-sphere}(b) and (c). 
Therefore, these configurations give maps from 
the space ${\mathbb R} \times S^1$ 
to a half of the target space.
BPS configurations 
$\omega_L$ and $\omega_R$ 
carry a half of the unit instanton charge, 
$Q= 1/2$, 
while anti-BPS configurations
$\omega_L^*$ and $\omega_R^*$ carry 
$Q=- 1/2$.
This fact also can be understood 
by noting that 
the $U(1)$ modulus is twisted half 
along the domain wall \cite{Nitta:2012xq,Nitta:2012kj}.

Fractionalized instantons can exist in the 
${\mathbb C}P^{N-1}$ model too. 
The configuration (\ref{eq:fractional}) of the 
${\mathbb C}P^{1}$ model can be generalized 
into the $N$-vector $\omega$ for the  
${\mathbb C}P^{N-1}$ model 
with the ${\mathbb Z}_N$ twisted boundary condition in Eq.~(\ref{ZNC}) as
\begin{eqnarray}
&& \omega_L
 = \left(0, \cdots, 0,1,
\lambda e^{i\theta}e^{+2\pi z/N} , 0,\cdot\cdot\cdot \right)^{T}\,, \quad
 \omega_R
 = \left(0,\cdots, 0,1,
\lambda e^{i\theta}e^{-2\pi z/N} ,0,\cdot\cdot\cdot,0\right)^{T}\,.\quad
\end{eqnarray}

\subsection{Neutral bions}
\begin{figure}[htbp]
\begin{center}
 \includegraphics[width=0.5\textwidth]{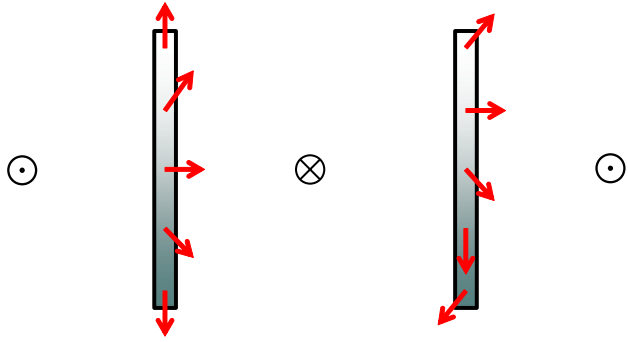}
\end{center}
\caption{Neutral bion.
This is a composite of fractionalized instantons 
$\omega_L$ and $\omega_R^*$, 
where we have introduced a relative phase.
The notation is the same as 
Fig.~\ref{fig:fractional-config}
\label{fig:bion-config}
}
\end{figure}

\begin{figure}[htbp]
\begin{center}
 \includegraphics[width=0.99\textwidth]{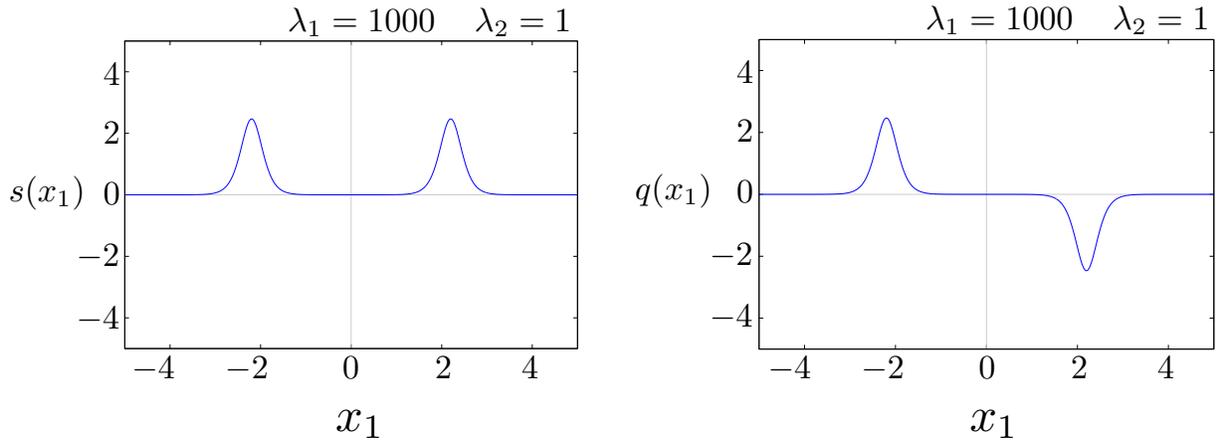}
\end{center}
\caption{Action density $s(x_{1})$ and topological charge 
density $q(x_{1})$ for the configuration of Eq.~(\ref{BBozero}) 
for $\lambda_{1}=1000$, $\lambda_{2}=1$ and $\theta_{2}=0$. 
The distance between the peaks of two fractionalized instantons 
is given by $\sim4.3976$, which is consistent with the 
separation $(1/\pi)\log(1000^{2})$ obtained from Eq.~(\ref{sepa}).
}
\label{BB1}
\end{figure}

A neutral bion configuration is 
a composite of a fractionalized instanton and 
fractionalized anti-intanton with 
the total instanton charge canceled out. 
Let us discuss the ${\mathbb C}P^1$ model first. 
From the solutions in Eq.~(\ref{eq:fractional}) 
and their complex conjugates, 
it is reasonable to consider
the following ansatz for  the ${\mathbb C}P^{1}$ model 
satisfying a ${\mathbb Z}_{2}$ twisted boundary condition (\ref{ZNC}) as 
\begin{equation}
\omega
 = \left(1+\lambda_{2}e^{i\theta_{2}}e^{\pi(z+\bar{z})},\,\,\lambda_{1}e^{i\theta_{1}}e^{\pi z} \right)^{T}\,, 
 \label{BBozero}
\end{equation}
constructed from fractionalized instantons $\omega_L$ and $\omega_R^*$ in Eq.~(\ref{eq:fractional}).  
As we mentioned, the ansatz $\omega=e^{-\pi z/2}\left(1+\lambda_{2}e^{i\theta_{2}}e^{\pi(z+\bar{z})},\,\,\lambda_{1}e^{i\theta_{1}}e^{\pi z} \right)^{T}$ also gives the same ${\bf P}(x)$, thus these are equivalent.
$\lambda_{1}\geq0$, $\lambda_{2}\geq0$, $0\leq\theta_{1} ,\theta_{2}< 2\pi$ 
are all real parameters characterizing the configuration 
associated with this ansatz, 
as $\lambda_{1}^{2}/\lambda_{2}$ and $\lambda_{2}$ govern a 
relative separation and a center location between the instanton 
constituents respectively.
We have no parameter characterizing the size
of fractionalized instantons in the present ansatz.
For $\lambda_{1}^{2}\gg \lambda_{2}$, this configuration is composed of two components, a BPS fractionalized instanton ($S=1/2$, $Q=1/2$) and a BPS fractionalized anti-instanton ($S=1/2$, $Q=-1/2$), 
which are separately located 
as shown in Fig.~\ref{fig:bion-config}.
Fig.~\ref{BB1} shows the action and topological charge 
densities of this configuration.

The superposition ansatz such as ours 
has been studied long for Yang-Mills instantons
and ${\mathbb C}P^{N-1}$ instantons on $S^{2}$ and ${\mathbb  R}^{2}$
(See \cite{lecture} for example).
On the other hand, for these theories, 
a multiple-type ansatz has been also investigated \cite{MD1}, 
in relation to the study on ``zindons".
However, due to the fixed twisted boundary condition, 
it is not straightforward to construct an ansatz
for the present case, with keeping, non-triviality of configurations, finite energy 
and the boundary conditions. 
The twisted boundary condition strongly restricts
patterns of ansatz. 
This is why we begin with the simple ansatz (\ref{BBozero}).

It is notable that the action density and topological charge 
density are independent of $\theta_{1}$. 
The operators $\bf{P}$, $\partial_{z}{\bf P}$, 
$\partial_{\bar{z}}{\bf P}$, 
$\partial_{z}{\bf P}\partial_{\bar{z}}{\bf P}$ and 
$\partial_{\bar{z}}{\bf P}\partial_{z}{\bf P}$ have the 
following form as
\begin{equation}
\left(
    \begin{array}{cc}
      a & b\,e^{-i\theta_{1}} \\
      c\,e^{+i\theta_{1}} & d
    \end{array}
  \right)\,,
\end{equation}
where $a,b,c$ and $d$ are some functions of $z$ and $\bar{z}$ including $\lambda_{1},\lambda_{2}$ and $\theta_{2}$ as parameters.
Then, it is obvious that both $s(x_{1})\sim {\rm Tr}[\partial_{z}{\bf P}\partial_{\bar{z}}{\bf P}]$
and $q(x_{1})={\rm Tr}[{\bf P}(\partial_{\bar{z}}{\bf P}\partial_{z}{\bf P}-\partial_{z}{\bf P}\partial_{\bar{z}}{\bf P})]$ have no $\theta_{1}$ dependence.
It means that $\theta_{1}$ corresponds to a bosonic zero mode, which does not cost the configuration energy.
On the other hand, the configuration depends on $\theta_{2}$.
For now we assume $\theta_{2}=0$, and will
consider $\theta_{2}\not=0$ cases later.

The total action and the net topological charge in the large-separation 
limit are given by
\begin{equation}
S=1,\,\,\,\,Q=0\,,
\label{Ini}
\end{equation}
respectively. 
We note that the topological charge is zero for any values of 
separation, 
and this configuration corresponds to a topologically trivial vacuum.

Generalization of this configuration into the ${\mathbb C}P^{N-1}$ 
model is straightforward as 
\begin{equation}
\omega
 = \left(0,\cdot\cdot\cdot,0,1+\lambda_{2}e^{i\theta_{2}}e^{2\pi(z+\bar{z})/N},\,\,\lambda_{1}e^{i\theta_{1}}e^{2\pi z/N},0,\cdot\cdot\cdot, 0 \right)^{T}\,. 
 \label{BBozeroN}
\end{equation}
The corresponding configuration again has no $\theta_{1}$ dependence.
For $\lambda_{1}^{2}\gg \lambda_{2}$, this configuration 
corresponds to a $1/N$ instanton ($S=1/N$, $Q=1/N$) 
and a $1/N$ anti-instanton ($S=1/N$, $Q=-1/N$) at large separations. 
The total action and the net topological charge in this 
large-separation limit are given by
\begin{equation}
S=2/N,\,\,\,\,Q=0\,,
\label{Ini}
\end{equation}
respectively.

As $x_{1}$ varies from $-\infty$ to $\infty$, 
the normalized complex vector $n(x_{1})$ takes the following 
three different values, which we denote as 
$n_{1}, n_{2}, n_{3}$,
\begin{equation}
n_{1}=(1,0,\cdot\cdot\cdot,0)^{T}\,\,\,\to\,\,\, n_{2}=(0,1,\cdot\cdot\cdot,0)^{T}\,\,\,\to\,\,\, 
n_{3}=(1,0,\cdot\cdot\cdot,0)^{T}\,,
\end{equation}
for $\lambda_{1}^{2}>\lambda_{2}$. 
The above three domains are divided by two critical points 
corresponding to the locations of the two kinks. 
As shown in \cite{DU1}, 
the two affine co-roots $\alpha_{i}$ and $\alpha_{j}$, which correspond to
the two kinks (fractionalized instantons) in Fig.~\ref{BB1} are given by
\begin{align}
\alpha_{i} &= n_{2}-n_{1}\,,
\\
\alpha_{j} &= -(n_{3}-n_{2})\,,
\end{align}
which satisfies
\begin{equation}
n(x_{1}=\infty) = n(x_{1}=-\infty) + \alpha_{i}-\alpha_{j}\,.
\end{equation}
In the present case, $\alpha_{i}$ and $\alpha_{j}$ are identical, which we define as 
$\alpha_{i}=\alpha_{j}\equiv\alpha$. It is given by
\begin{equation}
\alpha\,=\, (0,1,\cdot\cdot\cdot,0)^{T}\,- \,(1,0,\cdot\cdot\cdot,0)^{T} 
\,=\, (-1,1,\cdot\cdot\cdot,0)^{T}\,.
\end{equation} 
We note that $\alpha_{i}\cdot\alpha_{j}=\alpha\cdot\alpha>0$ for this case.

\begin{figure}[htbp]
\begin{center}
\includegraphics[width=0.5\textwidth]{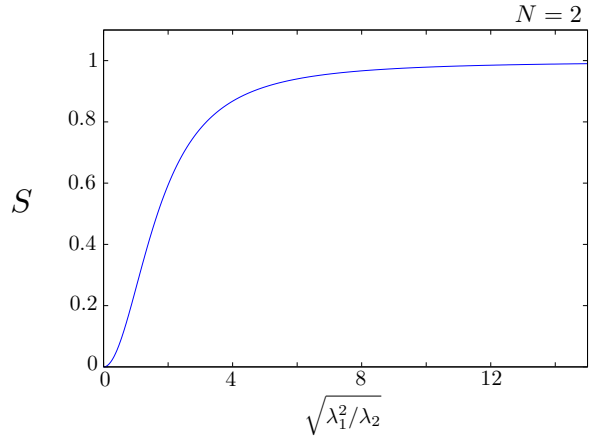}
\end{center}
\caption{The $\sqrt{\lambda_{1}^{2}/\lambda_{2}}$ dependence 
of the total action $S$ 
with $\theta_{2}=0$ for (\ref{BBozero}). 
The action is independent of $\lambda_{2}$ 
for $\lambda_{1}^{2}/\lambda_{2}$ fixed.
The configuration is changed from $S=1$ to $S=0$, due to the 
attractive force.}
\label{BBSF}
\end{figure}

The explicit form of the action density $s(x_{1})$ for general $N$ is given by
\begin{align}
s(x_{1})=
&{4\pi^{2}\over{N^{2}
\left(1+(\lambda_{1}^{2}+2\lambda_{2}\cos \theta_{2})e^{4\pi x_{1}/N}+\lambda_{2}^{2}e^{8\pi x_{1}/N}\right)^{4}}}\,\times
\nonumber\\
&\Big[
2(\lambda_{1}^{2}e^{4\pi x_{1}/N}-\lambda_{1}^{2}\lambda_{2}^{2}e^{12\pi x_{1}/N})^{2}
\nonumber\\
&
+(\lambda_{1}e^{2\pi x_{1}/N}+2\lambda_{1}\lambda_{2}e^{i\theta_{2}}e^{6\pi x_{1}/N}
+\lambda_{1}\lambda_{2}e^{i\theta_{2}}(\lambda_{1}^{2}+\lambda_{2}e^{i\theta_{2}})e^{10\pi x_{1}/N})\times
\nonumber\\
&\,\,\,\,\,\,\,
(\lambda_{1}e^{2\pi x_{1}/N}+2\lambda_{1}\lambda_{2}e^{-i\theta_{2}}e^{6\pi x_{1}/N}
+\lambda_{1}\lambda_{2}e^{-i\theta_{2}}(\lambda_{1}^{2}+\lambda_{2}e^{-i\theta_{2}})e^{10\pi x_{1}/N})
\nonumber\\
&
+(\lambda_{1}(\lambda_{1}^{2}+\lambda_{2}e^{i\theta_{2}})e^{6\pi x_{1}/N} +2\lambda_{1}\lambda_{2}^{2}
e^{10\pi x_{1}/N} +\lambda_{1}\lambda_{2}^{3}e^{-i\theta_{2}}e^{14\pi x_{1}/N} )\times
\nonumber\\
&\,\,\,\,\,\,\,
(\lambda_{1}(\lambda_{1}^{2}+\lambda_{2}e^{-i\theta_{2}})e^{6\pi x_{1}/N} +2\lambda_{1}\lambda_{2}^{2}
e^{10\pi x_{1}/N} +\lambda_{1}\lambda_{2}^{3}e^{i\theta_{2}}e^{14\pi x_{1}/N} )
\Big]\,.
\end{align}
Fig.~\ref{BBSF} depicts the $\sqrt{\lambda_{1}^{2}/\lambda_{2}}$ 
dependence of the total action $S$ with $\theta_{2}=0$ for $N=2$.

We will from now look into the attractive interaction between 
the two fractionalized instantons.
In order to understand precise separation dependence of action 
and interaction force,
we need to know the exact separation between the two components 
of fractionalized instantons in our configuration.
The positions $\tau_1$ and 
$\tau_2$ of fractionalized instantons 
and  fractionalized anti-instantons 
in the $x_1$-coordinate 
are given by the balance conditions 
\cite{Eto:2004rz,Isozumi:2004jc,Eto:2006mz,Eto:2006pg},
\begin{align}
&1=\lambda_{1} e^{2\pi\tau_{1}/N}\,\,\,\,\,\,\,\,\,\to\,\,\,\,\,\,\,\,\,\tau_{1}={N\over{2\pi}}\log \left({1\over{\lambda_{1}}}\right)\,,
\\
&\lambda_{2}e^{4\pi\tau_{2}/N}=\lambda_{1} e^{2\pi\tau_{2}/N}\,\,\,\,\,\,\,\,\,\to\,\,\,\,\,\,\,\,\,
\tau_{2}={N\over{2\pi}}\log \left({\lambda_{1}\over{\lambda_{2}}}\right)\,,
\end{align}
respectively. 
Then, the separation $\tau$ between them 
is given by
\begin{equation}
\tau=\tau_{2}-\tau_{1}={N\over{2\pi}}\log \left({\lambda_{1}^{2}\over{\lambda_{2}}}\right)\,.
\label{sepa}
\end{equation} 
For $\tau\geq0$, $\tau$ can be interpreted as separation 
between the fractionalized-instanton components.
The definition of separation depends on $N$ for this configuration.
As noted in the caption in Fig.~\ref{BB1}, 
this definition of separation precisely describes
the distance between the locations of two fractionalized instantons.

\begin{figure}[htbp]
\begin{center}
 \includegraphics[width=0.99\textwidth]{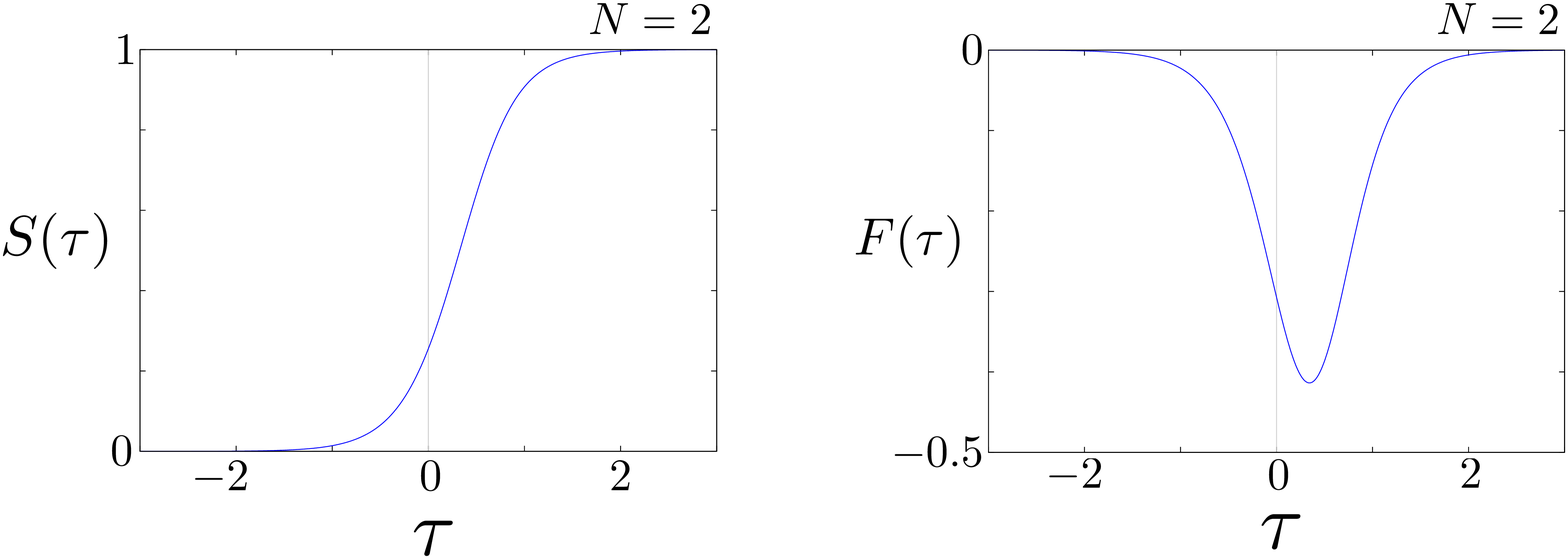}
\end{center}
\caption{The $\tau=(1/\pi)\log\lambda_{1}^{2}/\lambda_{2}$ 
dependence of the total action $S$ 
and the force $F=-{dS\over{d\tau}}$ with $\theta_{2}=0$ for 
(\ref{BBozero}).
For $\tau\geq0$, we can interpret $\tau$ as separation between 
the instanton constituents.
The configuration is changed from $S=1$ to $S=0$, due to the 
attractive force.
The configuration for $\tau\agt 1$ corresponds to neutral bions.}
\label{BBSFlog}
\end{figure}

Fig.~\ref{BBSFlog} depicts the separation $\tau$ dependence of the total action $S$
and the static force $F=-{dS\over{d\tau}}$ with $\lambda_{2}=1$ fixed for $N=2$.
It indicates that the total action monotonically decreases as $\tau$ gets smaller,
and the interaction force is negative for wide $\tau$ range.
It clearly shows that the fractionalized-instanton constituents 
exert an attractive force. 
To be precise, as will be shown later, the interaction force is exponentially 
suppressed for large separation $\tau \gg 1$ or the merged limit 
$\tau\ll 0$ ($\lambda_{1}^{2}/\lambda_{2}\ll1$).
It indicates that our ansatz yields intermediate configurations between
two (approximate) solutions, a two-separated fractionalized-instanton solution ($S=1$, $Q=0$)
and a trivial perturbative vacuum ($S=0$, $Q=0$).
Our analysis is easily generalized to $\lambda_{2}\not=1$, where we
find that the total action is independent of $\lambda_{2}$ 
if $\lambda_{1}^{2}/\lambda_{2}$ or $\tau$ are fixed.
From this analysis, we see that the location of the center of mass 
$\lambda_{2}$ corresponds to a bosonic zero mode 
while $\lambda_{1}^{2}/\lambda_{2}$ to a negative mode.

\begin{figure}[htbp]
\begin{center}
 \includegraphics[width=0.99\textwidth]{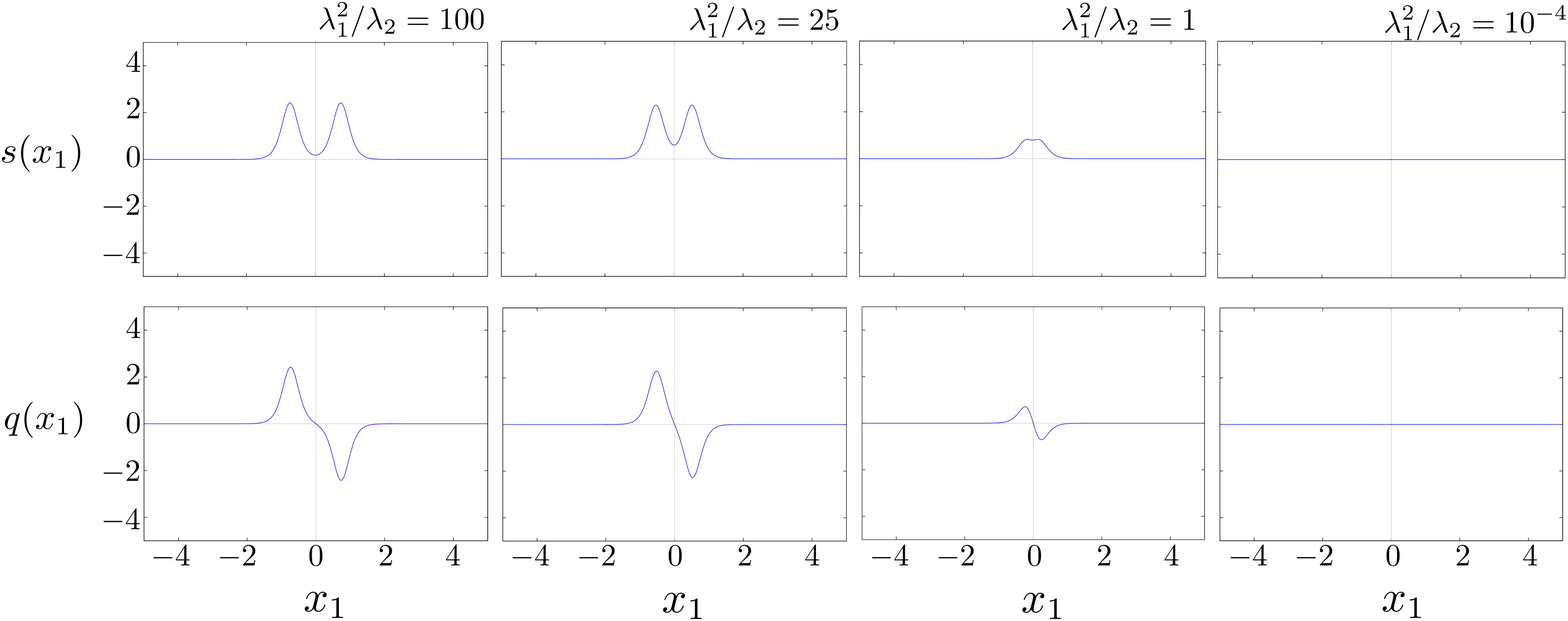}
\end{center}
\caption{Action density $s(x_{1})$ (up) and charge density 
$q(x_{1})$ (down) of the configuration of Eq.~(\ref{BBozero}) 
for $\lambda_{1}^{2}/\lambda_{2}=100, 25, 1, 10^{-4}$ 
($\tau=1.47, 1.02, 0, -2.93$) for $\lambda_{2}=1$ and $\theta_{2}=0$.
The configurations for $\lambda_{1}^{2}/\lambda_{2}=100, 25$ correspond
to neutral bions.
}
\label{BB2-1}
\end{figure}

The two constituents are getting closer and finally 
are merged by the attractive force, as shown in Fig.~\ref{BB2-1}.
The resultant configuration at $\tau=-\infty$ 
($\lambda_{1}^{2}/\lambda_{2}=0$) is given by
\begin{equation}
\omega(\tau =- \infty)
\, \to\, \left(1+\lambda_{2}e^{\pi(z+\bar{z})},\,\,0\right)^{T}\,, 
 \label{of}
\end{equation}
for $N=2$, and 
\begin{equation}
\omega(\tau =- \infty)
\, \to\, \left(0,\cdot\cdot\cdot,0,1+\lambda_{2}e^{i\theta_{2}}e^{2\pi(z+\bar{z})/N},\,\,0,0,\cdot\cdot\cdot, 0 \right)^{T}\,. 
\end{equation}
for general $N$, with the quantum number 
\begin{equation}  
S=0,\,\,\,\,Q=0\,,
\end{equation}
which is identical to a trivial vacuum. 

Here we discuss the characteristic size of neutral bions.
As shown in \cite{DU1}, the size of ``charged" bions $N_{f}>0$ is clearly
determined since the instanton constituents are bound due to
the balance of bosonic repulsive and fermionic attractive forces.
This calculation can be extended to the case of $N_{f}=0$, and it
gives the length scale of charged bions
as $\tau^{\ast}={N\over{2\pi}}\log {8\pi\over{g^{2}N}}$, 
which reads $\tau^{\ast}=0.5\sim1.5$ for $N=O(1)$ with $g^{2}=1$.
On the other hand, the calculation of the neutral bion size is not straightforward 
since its bosonic interaction is attractive and it requires analytic continuation of $g^{2}$
to negative values and its returning back to positive values.
However, taking into account the fact that the actions of neutral and charged bions in the amplitudes are common except for the imaginary part, 
we speculate that neutral bions have a similar 
size or length scale to that of the charged bions $\tau^{\ast} \sim 1$.    
In the rest of the paper, 
we assume that the neutral bions arise from the separation scale $\tau\agt1$.
We thus regard our ansatz as the neutral bions only for the separation $\tau>1$.
We will discuss whether this assumption is appropriate or not in terms of
BZJ-prescription in Sec.~\ref{sec:SD}.

\begin{figure}[htbp]
\begin{center}
 \includegraphics[width=0.99\textwidth]{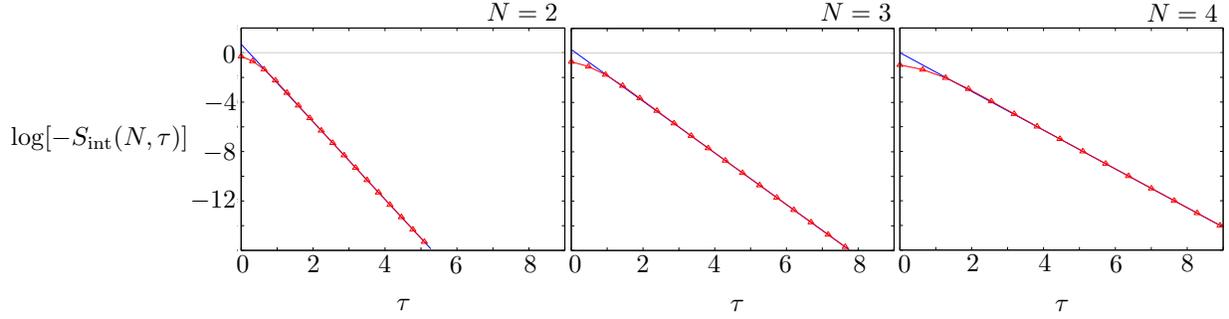}
\end{center}
\caption{Plot of $\log (-S_{\rm int}(N,\tau))$ as a function of $\tau$ 
for $N=2$ (left), $N=3$ (center) and $N=4$ (right) for (\ref{BBozeroN}) (red curves with 
triangle points).
For $\tau>1$, the curve is almost equivalent to $-(2\pi/N)\tau+ C(N)$ (blue curves).}
\label{Sint}
\end{figure}

We now investigate interaction part of the action for 
this configuration, to compare our concrete ansatz to
the far-separated instanton argument (\ref{bionS}) in Ref.~\cite{DU1}.
The interaction part of the action density is written 
as the action density $s(x_{1})$ minus
the one fractionalized-instanton density and one fractionalized-anti-instanton density  $s_{\nu=1/N}(x_{1})+s_{\nu=-1/N}(x_{1})$,
\begin{equation}
s_{\rm int}(x_{1}) = s(x_{1})-(s_{\nu=1/N}(x_{1})+s_{\nu=-1/N}(x_{1}))\,.
\end{equation}
The integrated interaction action is then given by
\begin{equation}
S_{\rm int}(N, \tau)\, =\, {1\over{ \pi}}\int dx\, s_{\rm int}(x_{1})\,.
\end{equation}
In Fig.~\ref{Sint}, we plot the logarithm of the total 
interaction action $S_{\rm int}(N, \tau)$ as a function of $\tau$
for $N=2,3,4$.
For $\tau\agt1$ region, $\log (-S_{\rm int}(N,\tau))$ can be 
well approximated by analytic lines,
\begin{equation}
\log \left[-S_{\rm int}(N,\tau)\right]\,\,\sim\,\, -\xi(N)\, \tau \,\,+\,\, C(N)\,,\,\,\,\,\,\,\,\,\,\,\,\,\,\,\,\,\, (\tau\agt1)\,,
\end{equation}
where $\xi(N)$ is a slope and $C(N)$ is a $y$-intercept.
In Fig.~\ref{Sint} we simultaneously depict these analytic lines for the three cases. 
The slopes $\xi$ of the approximate lines 
read $\xi\sim \pi$ for $N=2$, $\xi\sim 2\pi/3$ for $N=3$ 
and $\xi\sim \pi/2$ for $N=4$,
which indicates that the slope $\xi$ can be generally expressed as
\begin{equation}
\xi(N)\,\sim\, {2\pi\over{N}}\,.
\end{equation}
Therefore we observe 
that the interaction action can be written as the following 
form for $\tau\agt1$ region,
\begin{equation}
S_{\rm int} (N, \tau)\, \,\sim\,\, -\,e^{C} \, e^{-\xi \tau}\,, \,\,\,\,\,\,\,\,\,\,\,\xi={2\pi\over{N}}\,,
\,\,\,\,\,\,\,\,\,\,\,\,\,\,\,\,\, (\tau\agt1)\,.
\label{eq:int_action_approx}
\end{equation}
This $\xi$ is equivalent to the (dimensionless) lowest Kaluza-Klein spectrum $Lm_{LKK}$,
which is given as $Lm_{LKK} = |q_{i}-q_{j}|=2\pi/N$, where $q_{i}$ and $q_{j}$
are two nonzero components of Wilson-loop holonomy in (\ref{WHN})(\ref{WHN2}).

\begin{figure}[htbp]
\begin{center}
 \includegraphics[width=0.6\textwidth]{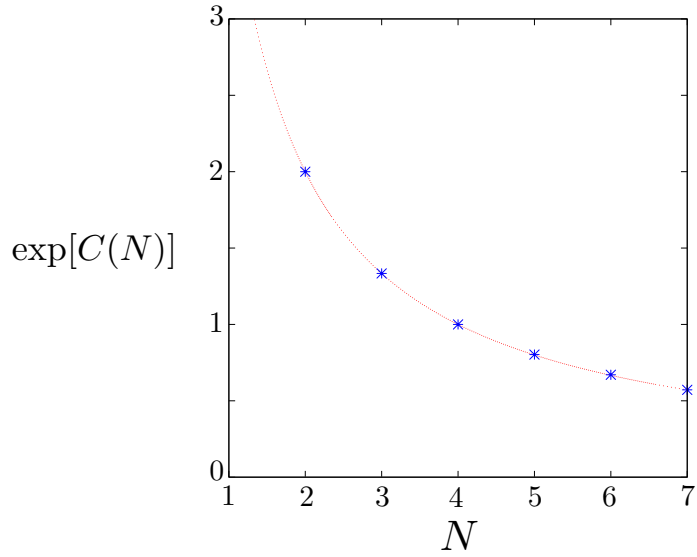}
\end{center}
\caption{The coefficient of the interaction action $\exp[C(N)]$ 
in Eq.(\ref{eq:int_action_approx}) 
as a function of $N$ for $N = 2,3,4,5,6,7$ for the 
Ansatz (\ref{BBozeroN}) (blue points). 
The coefficient can be approximated by $4/N$ (a red curve).}
\label{N-dep}
\end{figure}

We next determine the $N$ dependence of the $y$-intercept $C(N)$.
In Fig.~\ref{N-dep} we plot exponential of the intercept $\exp[{C(N)}]$ 
as a function of $N$ for $N=2,3,4,5,6,7$.
We find out that this dependence is well approximated by $\exp[C(N)]\sim4/N$, 
and depict it simultaneously in the figure.
This result shows that the interaction action for $\tau\agt1$ can be written as
\begin{equation}
S_{\rm int} (N, \tau)\,\,\sim\,\, -{4\over{N}} \, e^{-\xi \tau}\,,  \,\,\,\,\,\,\,\,\,\,\,\xi={2\pi\over{N}}\,,
\,\,\,\,\,\,\,\,\,\,\,\,\,\,\,\,\, (\tau\agt1)\,.
\end{equation}
It means, for a wide range of separations $\tau\agt1$,
the interaction part of the action for our configuration 
is consistent with the neutral bion action (\ref{bionS}) obtained from 
the far-separated instanton calculation up to $2\pi$ factor, 
which we introduced for convenience, as
\begin{equation}
S_{\rm int}(N, \tau)=-4\xi \,(\alpha_{i}\cdot \alpha_{j})\,e^{-\xi\tau} 
= -{8\pi\over{N}}e^{-\xi\tau}\,, \,\,\,\,\,\,\,\,\,\,\,\xi={2\pi\over{N}}\,,
\end{equation}
with $\alpha_{i}\cdot\alpha_{j}=\alpha\cdot\alpha = 1$ for our ansatz
following the Lie algebra notation in \cite{DU1}.

We have shown that our ansatz (\ref{BBozero}) gives a configuration 
consistent to (\ref{bionS}) except in the merged region $\tau<1$.
It means that (\ref{BBozero}) is a good ansatz describing the neutral bion,
and can be identified as an infrared renormalon since the imaginary part of its amplitude
obtained through the BZJ-prescription ($g^{2} \to -g^{2}$) and 
analytic continuation cancels the notorious ambiguity arising 
in the Borel re-summation of the perturbative series.
By use of the present ansatz, we can study properties of bions and the related physics,
not only at large separation $\tau\gg1$, but also at short separation $\tau\agt1$.

We here discuss a physical meaning of our ansatz for the merged region $\tau<1$.
The coincidence of the interactive actions for $\tau \agt 1$ (Fig.~\ref{Sint}) implies 
that the neutral bion scale can be determined by such a coincidence point,
or $\tau \sim 1$ for this case, which is consistent with the charged bion scale.
On the other hand, for the region $\tau<1$, 
the configurations (see the right two columns in Fig.~\ref{BB2-1}) 
are regarded as those around the perturbative vacuum rather than the bion saddle.
It means that, in the semi-classical calculation, these configurations correspond to
fluctuations around the perturbative saddle, but not around the bion saddle point. 
It is also notable that our ansatz connects these 
two different saddle points continuously by a single parameter.
We will briefly discuss how to classify the parameter regions into the two sectors
in Sec.~\ref{sec:SD}.

\begin{figure}[htbp]
\begin{center}
 \includegraphics[width=0.99\textwidth]{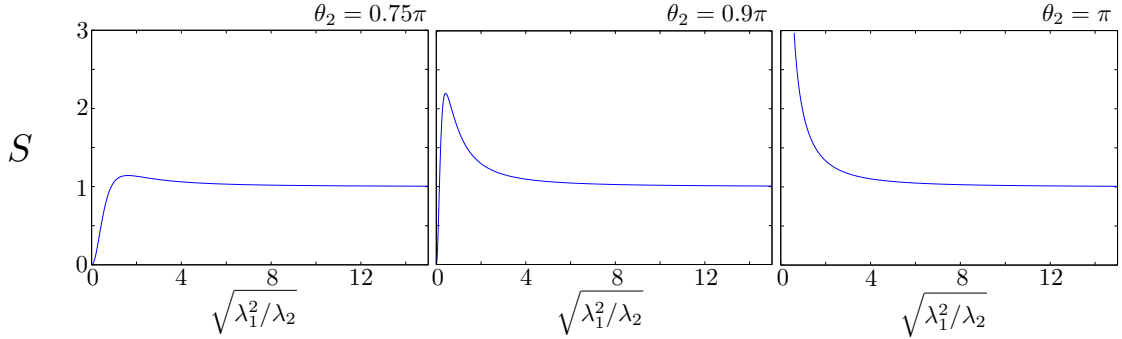}
\end{center}
\caption{The $\sqrt{\lambda_{1}^{2}/\lambda_{2}}$ dependence 
of the total action $S$ for the configuration in Eq.~(\ref{BBozero}) 
for $\theta_{2}=0.75\pi, 0.9\pi, \pi$.}
\label{bar}
\end{figure}

We here make a comment on cases for $\theta_{2}\not=0$.
For $0<\theta_{2}<\pi/2$, the interaction force is qualitatively 
the same as the case for $\theta_{2}=0$, or attractive.
For $\pi/2 \leq \theta_{2}\leq \pi$, things change:
The potential barrier emerges around 
$\sqrt{\lambda_{1}^{2}/\lambda_{2}}=1$ ($\tau=0$),
and the height becomes infinite for $\theta_{2}=\pi$ as shown 
in Fig.~\ref{bar}.
Of course, this does not mean that the interaction is repulsive since
$\theta_{2}$ is also a dynamical field variable and should 
relax eventually to 
$\theta_{2}=0$
in order to minimize the total action. 
The result indicates that $\theta_{2}$ corresponds to a positive mode.


\section{Bions with Non-${\mathbb Z}_{N}$ twisted boundary conditions}
\label{sec:oTBC}

The configuration discussed in the previous section is specific 
to the ${\mathbb Z}_{N}$ twisted boundary condition. 
In this section we consider different boundary conditions 
and bion-like configurations.
We here begin with the ${\mathbb C}P^{2}$ model and extend the 
result to the ${\mathbb C}P^{N-1}$. 

We first consider the following twisted boundary condition for 
${\mathbb C}P^{2}$
\begin{equation}
\omega(x_{1}, x_{2}+L) = \Omega \,\omega(x_{1},x_{2})\,,
\,\,\,\,\,\,\,\,\,\,\,\,\,
\Omega={\rm diag.}\left[e^{\pi i} ,  e^{\pi i} ,  1 \right]
={\rm diag.}\left[1 ,  1 ,  e^{-\pi i} \right]e^{\pi i}\,.
\label{SpC}
\end{equation}
This boundary condition corresponds to the vacuum
\begin{equation}
\langle A_{2} \rangle = (\pi, \pi, 0 )\,,
\end{equation}
where we omit permutation copies.
In gauge theory, this boundary condition is realized by special 
Wilson-loop holonomy in the exotic gauge-broken phase in $SU(3)$ 
gauge theory with 
adjoint quarks \cite{H1, H2},
where the gauge symmetry is broken as $SU(3)\,\to\, SU(2)\times U(1)$.
In this vacuum two elements of the Wilson-loop holonomy have the same value, 
but the other has a different one.
This phase is called ``split phase", 
thus we term the above boundary condition a ``split" twisted 
boundary condition.
Although how a neutral bion works in the split vacuum has not 
yet been elucidated,
it is worth investigating bion-like configurations in this vacuum.

\subsection{Bions for the split twisted boundary condition}
\begin{figure}[htbp]
\begin{center}
 \includegraphics[width=0.99\textwidth]{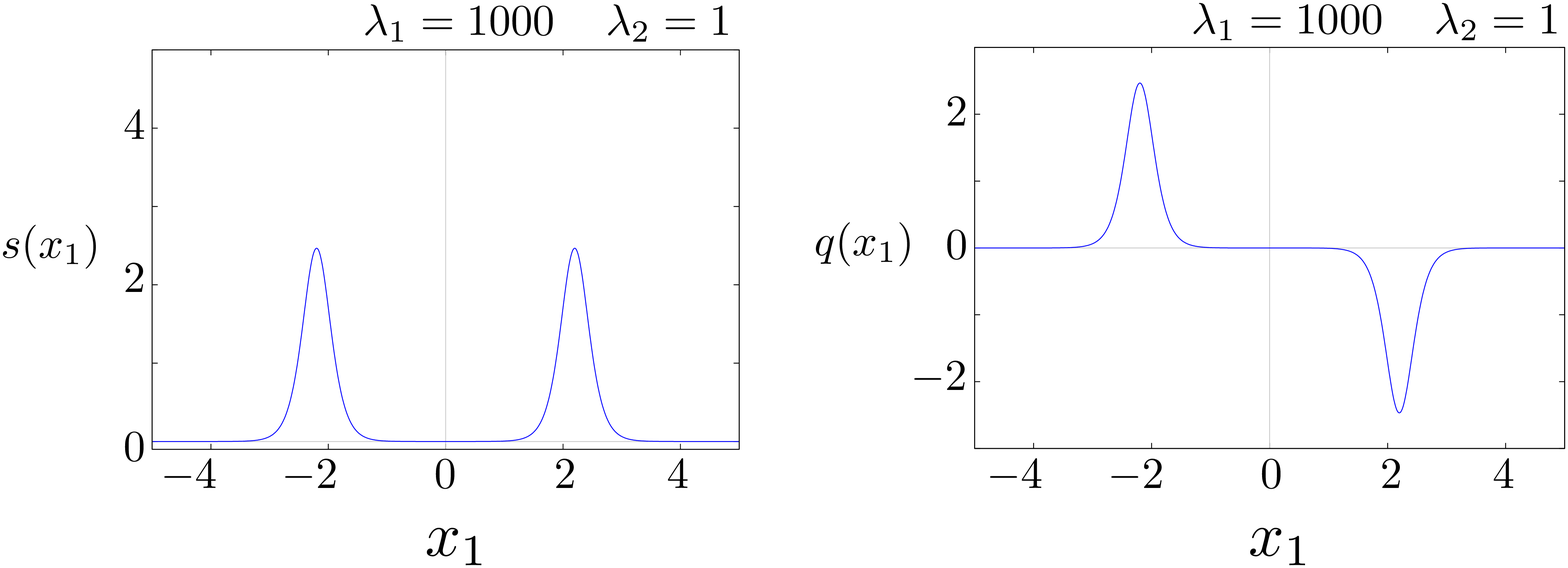}
\end{center}
\caption{Action density $s(x_{1})$ and topological charge 
density $q(x_{1})$ for 
the configuration in Eq.~(\ref{splitom}) for $\lambda_{1}=1000$ 
and $\lambda_{2}=1$. 
The distance between two fractionalized instantons is $\sim4.3976$, 
which is consistent 
with the $(1/\pi)\log (1000^{2})$ in Eq.~(\ref{sepa2}).
}
\label{Sp1}
\end{figure}

\begin{figure}[htbp]
\begin{center}
 \includegraphics[width=0.5\textwidth]{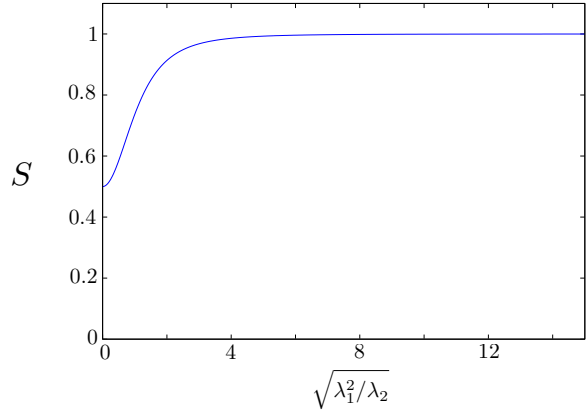}
\end{center}
\caption{The $\sqrt{\lambda_{1}^{2}/\lambda_{2}}$ dependence 
of the total action $S$ for the configuration in Eq.~(\ref{splitom}).
It is independent of $\lambda_{2}$ (corresponding to the center 
of two fractionalized instantons) 
with $\lambda_{1}^{2}/\lambda_{2}$ fixed.
The configuration is changed from $S=1$ to $S=1/2$,
due to the attractive force.}
\label{SpSF}
\end{figure}

We first consider a configuration in ${\mathbb C}P^{2}$ satisfying 
the split twisted boundary condition (\ref{SpC}),
\begin{equation}
\omega
 = \left(1,\,\,\lambda_{2} e^{i\theta_{2}}e^{\pi(z+\bar{z})},
\,\,\lambda_{1}e^{i\theta_{1}}e^{\pi z}\right)^{T}\,, 
 \label{splitom}
\end{equation}
This is an ansatz beyond the simple superposition ansatz.
For $\lambda_{1}^{2}\gg \lambda_{2}$, this configuration is composed 
of a BPS fractionalized instanton ($S=1/2$, $Q=1/2$) and a BPS 
fractionalized anti-instanton ($S=1/2$, $Q=-1/2$) in Fig.~\ref{Sp1}. 
The total action and the net topological charge in a far-separated 
limit are given by
\begin{equation}
S=1,\,\,\,\,Q=0\,.
\end{equation}
We note the action density and topological charge density are 
independent of the parameters $\theta_{1}$ and $\theta_{2}$.
Fig.~\ref{SpSF} depicts $\sqrt{\lambda_{1}^{2}/\lambda_{2}}$ 
dependence of the total action $S$ for (\ref{splitom}). 
The total action is independent of $\lambda_{2}$ with 
$\lambda_{1}^{2}/\lambda_{2}$ fixed.

For this ansatz, 
the normalized complex vector $n(x_{1})$ takes the following 
three different values as $x_{1}$ varies from $-\infty$ to $\infty$,
\begin{equation}
n_{1}=(1,0,0)^{T}\,\,\,\to\,\,\, n_{2}=(0,0,1)^{T}\,\,\,\to\,\,\, 
n_{3}=(0,1,0)^{T}\,,
\end{equation}
for $\lambda_{1}^{2}>\lambda_{2}$. 
For this case, the entry of extended Cartan matrix is again 
positive $\alpha_{i}\cdot\alpha_{j}>0$

\begin{figure}[htbp]
\begin{center}
 \includegraphics[width=0.99\textwidth]{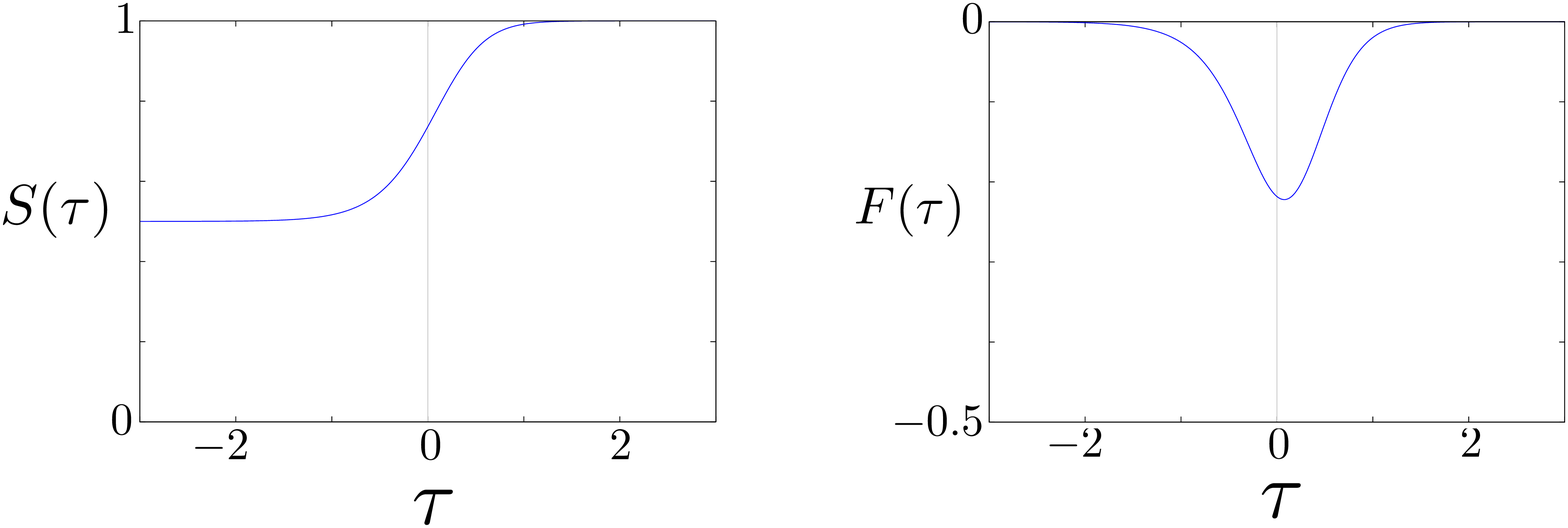}
\end{center}
\caption{The $\tau=(1/\pi)\log\lambda_{1}^{2}/\lambda_{2}$ 
dependence of the total action $S$ 
and the force $F=-{dS\over{d\tau}}$ for the configuration in 
Eq.~(\ref{splitom}).
For $\tau\geq0$, we can interpret $\tau$ as the separation 
between the instanton constituents.
The configuration is changed from $S=1$ to $S=1/2$ with $Q=0$ conserved,
due to the attractive force.
The configuration for $\tau\agt1$ corresponds to neutral bions.
}
\label{SpSFlog}
\end{figure}

The separation $\tau$ is given by
\begin{equation}
\tau={1\over{\pi}}\log \left({\lambda_{1}^{2}\over{\lambda_{2}}}\right)\,,
\label{sepa2}
\end{equation} 
which is obtained from the balance conditions
\cite{Eto:2004rz,Isozumi:2004jc,Eto:2006mz,Eto:2006pg},
\begin{align}
&1=\lambda_{1} e^{\pi\tau_{1}}\,\,\,\,\,\,\,\,\,\to\,\,\,\,\,\,\,\,\,
\tau_{1}={1\over{\pi}}\log \left({1\over{\lambda_{1}}}\right)\,,
\\
&\lambda_{2}e^{2\pi\tau_{2}}=\lambda_{1} e^{\pi\tau_{2}}\,\,\,\,\,\,\,\,\,\to\,\,\,\,\,\,\,\,\,
\tau_{2}={1\over{\pi}}\log \left({\lambda_{1}\over{\lambda_{2}}}\right)\,,
\end{align}
with $\tau=\tau_{2}-\tau_{1}$.
For $\tau\geq0$, $\tau$ stands for a separation between the 
fractionalized-instanton constituents.
In Fig.~\ref{SpSFlog} we depicts $\tau$ dependence of the total action $S$
and the static force $F=-{dS\over{d\tau}}$ for (\ref{splitom}).
The result indicates that the force is negative for $-\infty<\tau<\infty$, and 
the fractionalized instanton constituents have an attractive force. 
As with the configuration in the previous section, 
the two fractionalized instantons are merged by the attractive 
force, and finally resulting in the configuration,
\begin{equation}
\omega(\tau=-\infty)
\,\, \to\,\, \left(1,\,\,\lambda_{2} e^{i\theta_{2}}e^{\pi(z+\bar{z})},\,\,0\right)^{T}\,, 
\end{equation}
 at $\tau=-\infty$ ($\lambda_{1}^{2}/\lambda_{2}=0$) with 
\begin{equation}  
S=1/2,\,\,\,\,Q=0\,.
\end{equation}
From this analysis, we see that
$\lambda_{2}$, $\theta_{1}$ and $\theta_{2}$ correspond 
to bosonic zero modes 
while $\lambda_{1}^{2}/\lambda_{2}$ to a negative mode.

\begin{figure}[htbp]
\begin{center}
 \includegraphics[width=0.8\textwidth]{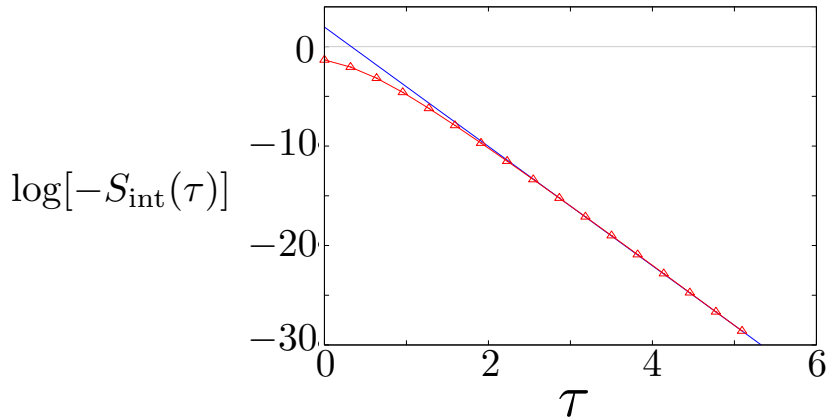}
\end{center}
\caption{Plot of $\log (-S_{\rm int}(\tau))$ as a function of $\tau$ 
for (\ref{splitom}) (a red curve with triangle points).
For $\tau>2$, the curve is almost equivalent to $-6\tau+ 1.97$ (a blue curve).}
\label{tsp_N=3}
\end{figure}

The interaction part of the action takes the same form as that in the previous section
$S_{\rm int}(\tau)= {1\over{ \pi}}\int dx s_{\rm int}(x_{1})$.
In Fig.~\ref{tsp_N=3}, we plot the logarithm of the interaction action 
$S_{\rm int}(\tau)$ as a function of $\tau$ for the present case (\ref{splitom}).
For $\tau\agt2$ region, $\log (-S_{\rm int}(\tau))$ is approximated by
\begin{equation}
\log \left[-S_{\rm int}(\tau)\right]\,\,\sim\,\, -6\, \tau \,\,+\,\, 1.9657\,,\,\,\,\,\,\,\,\,\,\,\,\,\,\,\,\,\, (\tau\agt2)\,.
\end{equation}
Therefore, the interaction action can be written as the following 
form for $\tau\agt2$ region,
\begin{equation}
S_{\rm int} (\tau)\, \,\sim\,\, -\,7.14 \, e^{-6 \tau}\,.
\,\,\,\,\,\,\,\,\,\,\,\,\,\,\,\,\, (\tau\agt2)\,.
\end{equation}
This is qualitatively consistent with (\ref{bionS}), while the coefficients are different from 
$\xi(3)$ and $C(3)$ in ${\mathbb Z}_{3}$ twisted boundary conditions.
Although how the coefficients are fixed needs to be uncovered,
we at least argue that neutral bion-type configurations exist also for the split boundary condition with $N=3$, which are responsible for the cancellation of infrared renormalon ambiguity. 

\subsection{Bions in extended split boundary conditions}

As an extension of the split boundary condition, 
we consider the following ansatz of the ${\mathbb C}P^{2}$ model 
on ${\mathbb R}^{1}\times S^{1}$,
\begin{equation}
\omega
 = \left(1,\,\,\lambda_{2} e^{i\theta_{2}}e^{{2\pi\over{3}}(z+\bar{z})},
\,\,\lambda_{1}e^{i\theta_{1}}e^{{2\pi\over{3}}z}\right)^{T}\,.
 \label{ozero}
\end{equation}
In this case we no longer regard the boundary condition as the 
Wilson-loop holonomy in 
the split vacuum, 
rather one specific twisted boundary condition, with $\Omega = (1,1,e^{2\pi i/3})$.
We will investigate the ansatz from pure-theoretical interest.

\begin{figure}[htbp]
\begin{center}
 \includegraphics[width=0.99\textwidth]{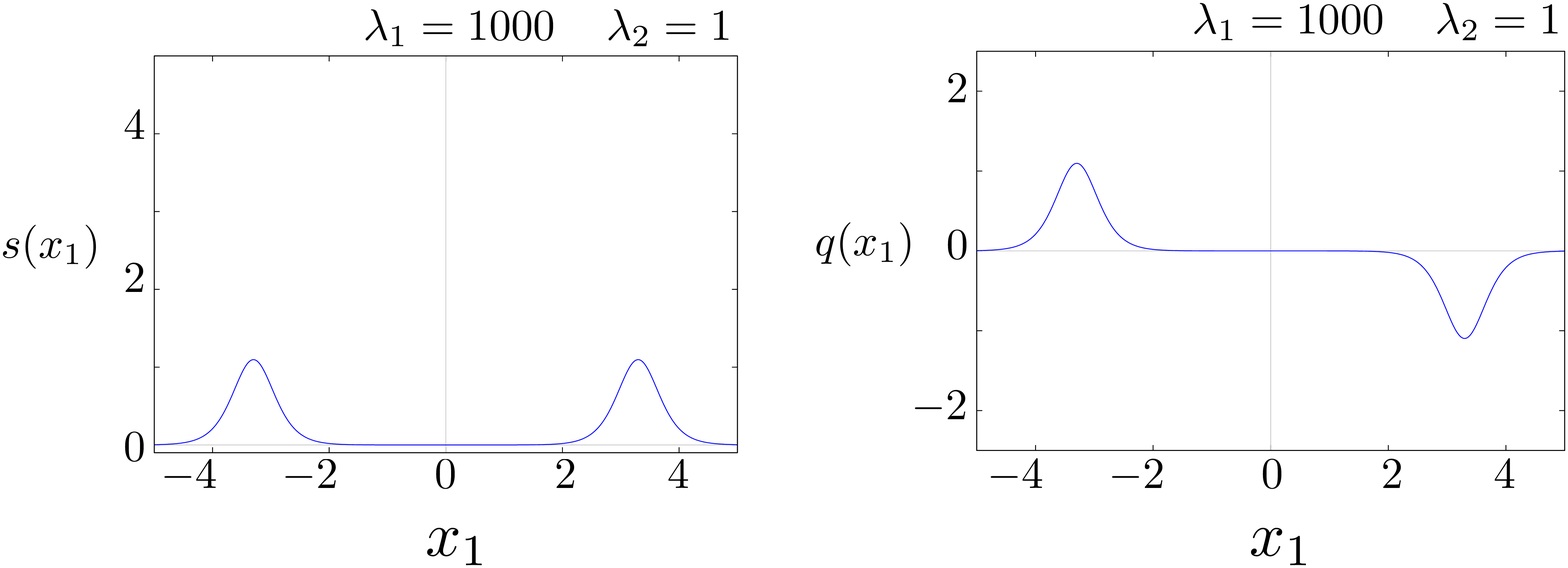}
\end{center}
\caption{Action density $s(x_{1})$ and topological charge 
density $q(x_{1})$ for the configuration of Eq.~(\ref{ozero}) 
for $\lambda_{1}=1000$ and $\lambda_{2}=1$.
The distance between two fractionalized instantons is $\sim6.596$, 
which is consistent with the $(3/2\pi)\log (1000^{2})$
in Eq.~(\ref{sepa3}).
}
\label{B1}
\end{figure}

For $\lambda_{1}^{2}\gg \lambda_{2}$, this configuration is 
composed of two constituents, a BPS fractionalized instanton 
($S=1/3$, $Q=1/3$) and a BPS fractionalized anti-instanton 
($S=1/3$, $Q=-1/3$), which are separately located as shown 
in Fig.~\ref{B1}. 
The total action and the net topological charge in this limit are given by
\begin{equation}
S=2/3,\,\,\,\,Q=0\,,
\label{Ini}
\end{equation}
respectively. 
We note that the topological charge is zero and conserved.

We generalize this configuration to the ${\mathbb C}P^{N-1}$ model,
\begin{equation}
\omega
 = \left(1,\,\,\lambda_{2}e^{i\theta_{2}}e^{{2\pi\over{N}}(z+\bar{z})},
 \,\,\lambda_{1}e^{i\theta_{1}}e^{{2\pi\over{N}}z},\,\,
 ...\,\,,0 \right)^{T}\,.
 \label{ozeroN}
\end{equation}
For $\lambda_{1}^{2}\gg \lambda_{2}$, this configuration corresponds to a $1/N$ instanton ($S=1/N$, $Q=1/N$) 
and a $1/N$ anti-instanton ($S=1/N$, $Q=-1/N$). 
The total action and the net topological charge in this 
large-separation limit are given by
\begin{equation}
S=2/N,\,\,\,\,Q=0\,,
\label{Ini}
\end{equation}
respectively. 
\begin{figure}[htbp]
\begin{center}
 \includegraphics[width=0.5\textwidth]{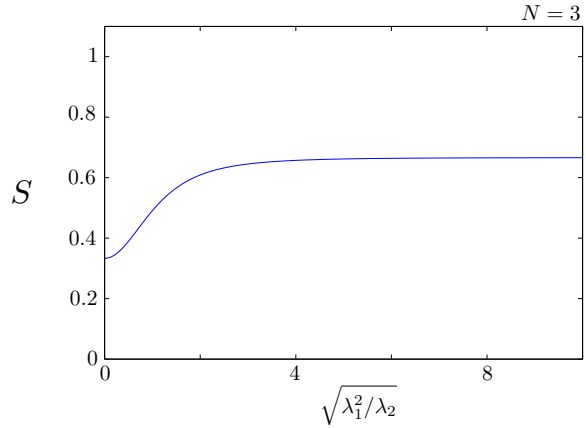}
\end{center}
\caption{The $\sqrt{\lambda_{1}^{2}/\lambda_{2}}$ dependence 
of the total action $S$ for the configuration in Eq.~(\ref{ozero}).
It is independent of the center location $\lambda_{2}$ with $\lambda_{1}^{2}/\lambda_{2}$ fixed.
The configuration is changed from $S=2/3$ to $S=1/3$ with $Q=0$ conserved,
due to the attractive force.}
\label{SF}
\end{figure}
The explicit form of the action density $s(x_{1})$ for $N$ is given by
\begin{align}
s(x_{1})=
&{4\pi^{2}\over{N^{2}(1+\lambda_{1}^{2}e^{4\pi x_{1}/N}+\lambda_{2}^{2}e^{8\pi x_{1}/N})^{4}}}\,\times
\nonumber\\
&\Big[
\lambda_{1}^{2}e^{4\pi x_{1}/N}(1+\lambda_{1}^{2}e^{4\pi x_{1}/N})^{2}
+2\lambda_{2}^{2}e^{8\pi x_{1}/N}(1+\lambda_{1}^{2}e^{8\pi x_{1}/N})^{2}
\nonumber\\
&+\lambda_{1}^{2}\lambda_{2}^{2}e^{12\pi x_{1}/N}(7+6\lambda_{1}^{2}e^{4\pi x_{1}/N}+7\lambda_{2}^{2}e^{8\pi x_{1}/N}
+2\lambda_{1}^{2}\lambda_{2}^{2}e^{12\pi x_{1}/N}+\lambda_{1}^{4}e^{8\pi x_{1}/N}+\lambda_{2}^{2}e^{16\pi x_{1}/N})
\Big]\,.
\end{align}
Fig.~\ref{SF} depicts the $\sqrt{\lambda_{1}^{2}/\lambda_{2}}$ 
dependence of the total action $S$ for $N=3$.
We note that the total action is independent of the central location $\lambda_{2}$
for $\lambda_{1}^{2}/\lambda_{2}$ fixed.
\begin{figure}[htbp]
\begin{center}
 \includegraphics[width=0.99\textwidth]{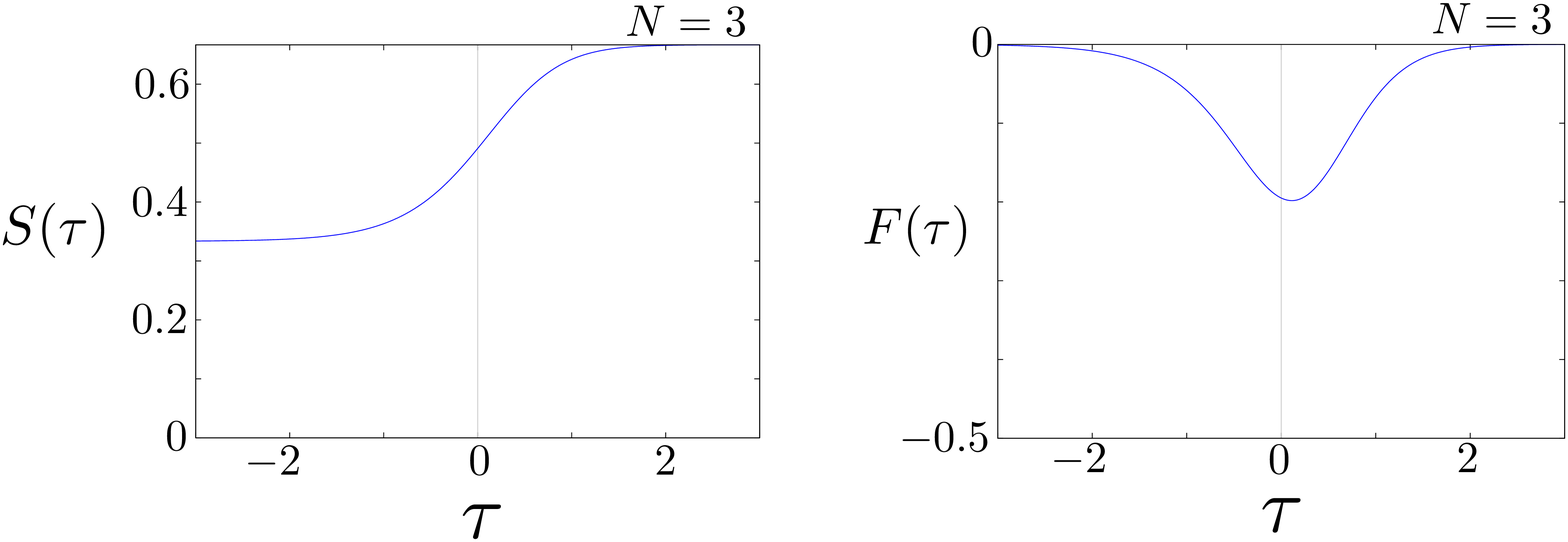}
\end{center}
\caption{The $\tau=(3/2\pi)\log\lambda_{1}^{2}/\lambda_{2}$ 
dependence of the total action $S$ 
and the force $F=-{dS\over{d\tau}}$ for the configuration in Eq.~(\ref{ozero}).
For $\tau\geq0$, we can interpret $\tau$ as the separation between 
the instanton constituents. 
The configuration is changed from $S=2/3$ to $S=1/3$ with $Q=0$ conserved, due to the attractive force.
The configuration for $\tau\agt1$ corresponds to neutral bions.
}
\label{SFlog3}
\end{figure}

The separation $\tau$ is given by
\begin{equation}
\tau={N\over{2\pi}}\log \left( {\lambda_{1}^{2}\over{\lambda_{2}}} \right)\,,
\label{sepa3}
\end{equation} 
which is obtained from the two balance conditions
\cite{Eto:2004rz,Isozumi:2004jc,Eto:2006mz,Eto:2006pg},
\begin{align}
&1=\lambda_{1} e^{2\pi\tau_{1}/N}\,\,\,\,\,\,\,\,\,\to\,\,\,\,\,\,\,\,\,\tau_{1}={N\over{2\pi}}\log \left({1\over{\lambda_{1}}}\right)\,,
\\
&\lambda_{2}e^{4\pi\tau_{2}/N}=\lambda_{1} e^{2\pi\tau_{2}/N}\,\,\,\,\,\,\,\,\,\to\,\,\,\,\,\,\,\,\,
\tau_{2}={N\over{2\pi}}\log \left({\lambda_{1}\over{\lambda_{2}}}\right)\,,
\end{align}
with $\tau=\tau_{2}-\tau_{1}$.
In Fig.~\ref{SFlog3} we depict the $\tau$ dependence of the total action $S$
and the static force $F=-{dS\over{d\tau}}$ for $N=3$.
The result clearly shows that the fractionalized instanton constituents have the attractive force. 

\begin{figure}[htbp]
\begin{center}
 \includegraphics[width=0.99\textwidth]{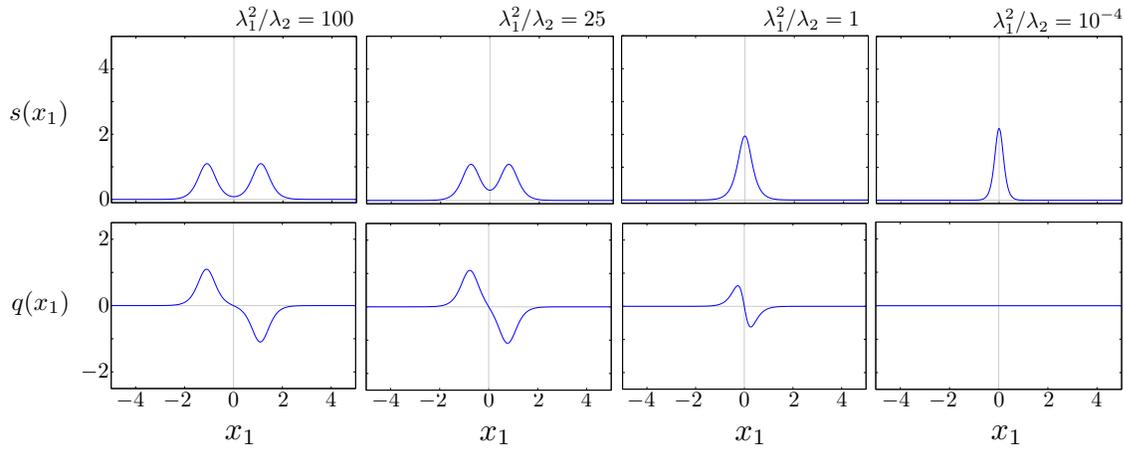}
\end{center}
\caption{Action density $s(x_{1})$ (up) and charge density $q(x_{1})$ (down) of
the configuration of Eq.~(\ref{ozero}) ($N=3$) 
for $\lambda_{1}^{2}/\lambda_{2}=100, 25, 1, 10^{-4}$ 
($\tau=2.20, 1.54, 0, -4.40$) with $\lambda_{2}=1$ fixed. 
The configurations for $\lambda_{1}^{2}/\lambda_{2}=100, 25$ correspond
to neutral bions.
}
\label{B2}
\end{figure}
The two constituents are merged 
by the attractive force, as shown in Fig.~\ref{B2}.
 For $N=3$, the configuration results in the following form 
\begin{equation}
\omega(\tau=-\infty)
 \,\,\to\,\, \left(1,\,\,\lambda_{2} 
e^{i\theta_{2}}e^{{2\pi\over{3}}(z+\bar{z})},\,\,0\right)^{T}\,, 
 \label{of}
\end{equation}
 at $\tau=-\infty$ ($\lambda_{1}^{2}/\lambda_{2}=0$) with
\begin{equation}  
S=1/3,\,\,\,\,Q=0\,.
\end{equation}
For general $N$, the resultant configuration at $\tau=-\infty$ ($\lambda_{1}^{2}/\lambda_{2}=0$) is given by
\begin{equation}
\omega(\tau=-\infty)
 \,\,\to\,\, \left(1,\,\,\lambda_{2} e^{i\theta_{2}}e^{{2\pi\over{N}}(z+\bar{z})}\,\,0,\,\,...\,\,0\right)^{T}\,, 
 \label{of}
\end{equation}
with
\begin{equation}  
S=1/N,\,\,\,\,Q=0\,.
\end{equation}

We now investigate the interaction part of 
the action for this configuration.
The interaction action is given by
\begin{equation}
S_{\rm int}(N, \tau) = {1\over{\pi}}\int dx s_{\rm int}(x_{1})\,,
\end{equation}
\begin{equation}
s_{\rm int}(x_{1}) = s(x_{1})-(s_{\nu=1/N}(x_{1})+s_{\nu=-1/N}(x_{1}))\,.
\end{equation}
\begin{figure}[htbp]
\begin{center}
 \includegraphics[width=0.99\textwidth]{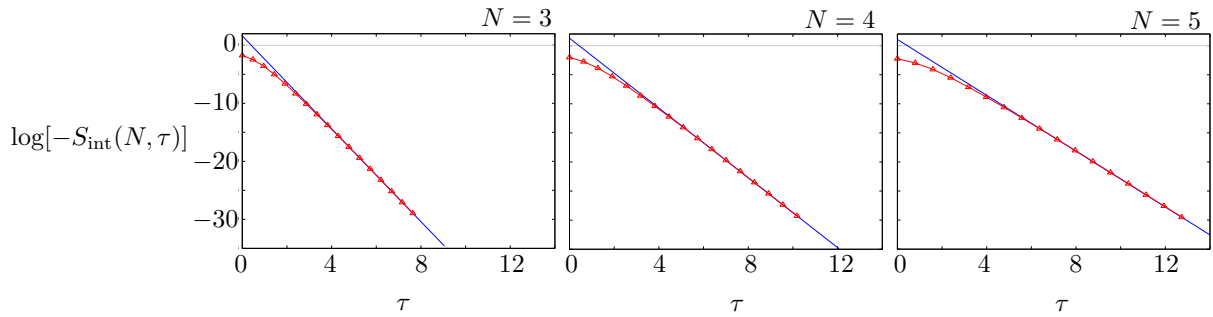}
\end{center}
\caption{Plot of $\log (-S_{\rm int}(N,\tau))$ as a function 
of $\tau$ for $N=3$ (left), $N=4$ (center) and $N=5$ (right) 
for the configuration in Eq.~(\ref{ozeroN}) (red curves with triangle points).
For sufficiently large $\tau$, the curve is approximated by
$-(12/N)\tau+ C(N)$ (blue curves).}
\label{sp_Sint}
\end{figure}
In Fig.~\ref{sp_Sint}, we plot the logarithm of the total 
interaction action $S_{\rm int}(N, \tau)$ as a function of $\tau$
for $N=2,3,4$.
For sufficiently large separation $\tau\agt 4$, 
$\log (-S_{\rm int})$ can be approximated by analytic lines,  
\begin{equation}
\log \left[-S_{\rm int}(N,\tau)\right]\,\,\sim\,\, -\xi(N)\, \tau \,\,+\,\, C(N)\,,
\,\,\,\,\,\,\,\,\,\,\,\,\,\,\,\,\,\,\,\,\,\,\,\,\,\,\,\,\, (\tau\agt4)\,,
\end{equation}
where $\xi(N)$ is a slope and $C(N)$ is an $y$-intercept.
For this case, the slope is expressed as
\begin{equation}
\xi(N) \sim {12\over{N}}\,.
\end{equation}
The interaction action can be written as the following form for large $\tau$ region,
\begin{equation}
S_{\rm int} (N, \tau)\, \,\sim\,\, -\,e^{C} \, e^{-\xi \tau}\,, \,\,\,\,\,\,\,\,\,\,\,\xi={12\over{N}}\,,
\,\,\,\,\,\,\,\,\,\,\,\,\,\,\,\,\, \,\,\,\,\,\,\,\,\,\,\,\,\,\,\,\,(\tau\agt4)\,.
\end{equation}
We next determine the $N$ dependence of $C(N)$.
\begin{figure}[htbp]
\begin{center}
 \includegraphics[width=0.6\textwidth]{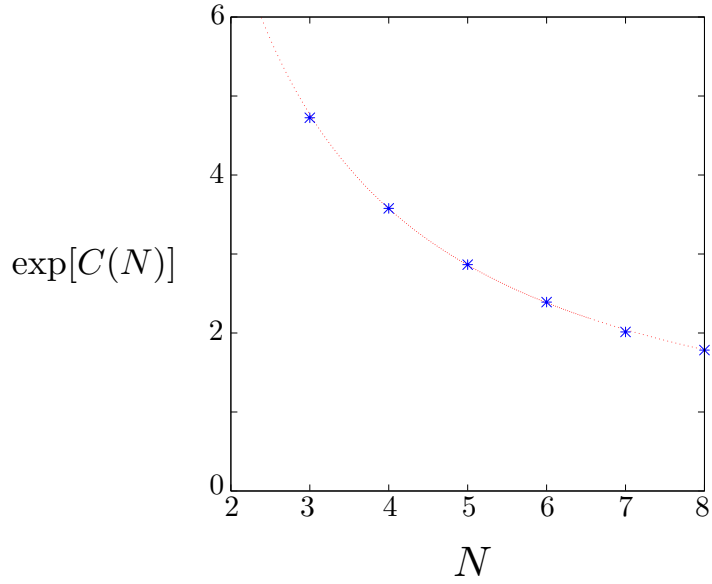}
\end{center}
\caption{Plot of $\exp[C(N)]$ (a coefficient of the interaction potential) 
for $N = 3,4,5,6,7,8$ for the configuration in Eq.~(\ref{ozeroN}) 
(blue points).
The plot is approximated by $14.3/N$ (a red curve).}
\label{sp_N-dep}
\end{figure}
In Fig.~\ref{sp_N-dep} we plot $\exp[C(N)]$ for $N=3,4,5,6,7,8$.
We find out that it is approximated by $\exp[C(N)]\sim14.3/N$, 
and depict it simultaneously in the figure.
This result indicates that the interaction action (potential) 
for large-separation region can be written as
\begin{equation}
S_{\rm int} (N, \tau) \sim -{14.3\over{N}} \, e^{-\xi \tau}\,, \,\,\,\,\,\,\,\,\,\,\,\xi={12\over{N}}\,,
\,\,\,\,\,\,\,\,\,\,\,\,\,\,\,\,\,\,\,\,\,\,\,\,\,\,\,\, (\tau\agt4)\,,
\end{equation}
which implies that the interaction part of the action at large 
$\tau$ region is expressed as
\begin{equation}
S_{\rm int}(\tau) \propto -\xi e^{-\xi\tau}\,.
\label{sp_bionS}
\end{equation}
This asymptotic form of the interaction potential is qualitatively
consistent to (\ref{bionS}).
For our special boundary conditions in the present subsection, 
which can no longer be identified as Wilson-loop holonomy, 
it is not straightforward to understand
the meaning of the values of $\xi$ and $e^{C}$. 
However, as with the case for the ${\mathbb Z}_{N}$ twisted 
boundary condition, it is true that the prescription 
($g^{2} \to -g^{2}$) and analytic continuation 
lead to the ambiguity in the imaginary part of the amplitude 
for this case too.
It implies that the resurgence procedure based on neutral 
bions universally 
works for general boundary conditions and vacua in field theories.

By calculating the renormalon ambiguity in the Borel re-summation 
of the perturbative series for the present non-${\mathbb Z}_{N}$ 
boundary conditions,
we can check if the two ambiguities are cancelled against each other.
In the future work, we will investigate whether resurgence procedure 
based on bions or bion-like configuratons still works 
for non-${\mathbb Z}_{N}$ vacuum such as split phases and its extensions.


\section{Summary and Discussion}
\label{sec:SD}

In this paper, we have revisited topologically trivial configurations 
in the ${\mathbb C}P^{N-1}$ model on ${\mathbb R}^{1}\times S^{1}$ 
with twisted boundary conditions, to study properties of 
bions composed of multiple fractionalized-instantons.
In the ${\mathbb C}P^{N-1}$ model with center-symmetric and 
non-center-symmetric twisted boundary conditions,
we have considered an explicit ansatz of a configuration containing 
one fractionalized instanton ($\nu=1/N$) and one fractionalized 
anti-instanton ($\nu=-1/N$), which has an attractive force.
We have shown that the separation-dependence and $N$-dependence of 
the interaction potential of the ansatz agree with the results 
of the far-separated instanton calculus \cite{DU1}, even at small values of the separation.

In Sec.~\ref{sec:ZN}, we have considered a simple neutral-bion 
ansatz for the ${\mathbb Z}_{N}$ twisted boundary condition, 
which represents a molecule of one fractionalized instanton ($\nu=1/N$) 
and one fractionalized anti-instanton ($\nu=-1/N$). 
From the separation dependence of the total action we show
that the interaction between the instanton constituents are 
attractive, thus the configuration has a negative mode.
The separation dependence and $N$-dependence of the interaction 
potential between the instanton constituents is compared with 
the result in the standard far-separated instanton calculus in 
Eq.~(\ref{bionS}), we show that our ansatz is consistent with 
Eq.~(\ref{bionS}) even from short to large separations.
This result indicates that our ansatz well describes
the neutral bion related to renormalon ambiguity,
which can be used from short ($\tau\agt 1$) to long ($\tau \gg 1$) separations.

In Sec.~\ref{sec:oTBC}, we have proposed bion-like ansatze
in non-${\mathbb Z}_{N}$ twisted boundary conditions including 
the one corresponding to the split vacuum in QCD(adj.) and 
its extensions for $N\geq3$.
We have shown that the interaction between the constituents 
is again attractive.
In this case, we have found that the separation and $N$ dependences 
of the interaction potential at large separation is qualitatively 
consistent to the result for ${\mathbb Z}_{N}$ twisted boundary conditions \cite{DU1} 
up to a numerical coefficient.
It implies that the bion resurgence procedure universally
works for a wide range of boundary conditions and vacua in field theories.

Our ansatz in the ${\mathbb C}P^{N-1}$ model corresponding to 
bion configurations can be a good starting point for studying 
properties of bions and related physics explicitly.
Indeed, by using our ansatz, we can study physics related to 
neutral bions, not only at large separations $\tau \gg 1$, 
but also at short separations $\tau\agt 1$, 
which cannot be reached by the far-separated instanton approach.

We here discuss how to determine 
which of the saddle points (perturbative or bion) 
configurations around $\tau\sim1$ should be classified into.
We so far have no systematic way to classify the parameters to the two associated regions 
by looking into our ansatz itself.
On the other hand, from the viewpoint of the BZJ-prescription,
the length scale where the imaginary ambiguity in the amplitude for our ansatz 
gets close to the ambiguity in the perturbative Borel-sum calculation should be
regarded as the neutral bion scale.
In our calculation, this scale is about $\tau \sim 1$ as shown in Fig.~\ref{Sint},
which is consistent with the charged bion scale too. 
Thus, for now, what we can do for this purpose is just to sort out the configurations
by the separation of the instanton constituents 
based on the plausible bion size $\tau\sim 1$, as performed in this paper.
Exact classification of the parameter regions for the two saddles should be
pursued in the future study.

As a future work, we consider to study ¡Ècharged bion¡É configurations, 
whose instanton constituents have a repulsive interaction in a bosonic sector 
and also have an attractive interaction due to the fermion zero mode exchange.
Due to the balance between the attractive and repulsive interactions, 
the size of charged bions will be clearly determined
and there will be no complicated problem on the size of bions for this case.
In the \"{U}nsal's argument \cite{U1, U2}, this configuration has great significance 
in weak-coupling-regime confinement via ``bion condensation".
While understanding of phase diagram in the $L$-$m_{adj}$ plane 
is required to elucidate its relation to confinement in pure Yang-Mills or QCD theories, 
it should be also worth investigating a concrete configuration 
contributing confinement in a toy model.

One straightforward extension will be 
bions in the Grassmanian sigma model 
with the target space $SU(N)/[SU(N-M)\times SU(M)\times U(1)]$. 
Domain walls in the Grassmannian sigma model 
were constructed in Ref.~\cite{Isozumi:2004jc,Eto:2005yh}.
Fractionalized instantons and bions can be composed from 
these solutions
with twisting $U(1)$ moduli. 
While the ${\mathbb C}P^1$ model with the twisted boundary condition 
has the Wilson-loop holonomy of  
a $U(1)$ gauge field,  the Grassmanian sigma model 
with the twisted boundary condition 
can have that of a non-Abelian gauge field.
We will see that the Grassmanian sigma model admits 
charged bions in addition to neutral bions. 
The D-brane configurations in Ref.~\cite{Eto:2006mz} 
will turn out to be very useful for analyzing this model.

One path to connect our results of bions in the
${\mathbb C}P^{N-1}$ model to QCD may be to consider 
a non-Abelian vortex \cite{Hanany:2003hp,Auzzi:2003fs,Eto:2005yh} 
in Yang-Mills theory in the Higgs vacuum.
$U(N)$ Yang-Mills theory coupled with suitable number of Higgs matter fields in the fundamental representation admits a non-Abelian vortex,
whose effective theory can be described by 
the ${\mathbb C}P^{N-1}$ model.
In this case, the Yang-Mills instantons and monopoles 
become ${\mathbb C}P^{N-1}$ instantons and domain walls, respectively,  
when trapped inside a vortex
\cite{Tong:2003pz,Shifman:2004dr,Hanany:2004ea,
Eto:2004rz,Eto:2006pg,Fujimori:2008ee}. 
Therefore, when the vortex world-sheet is wrapped around  $S^1$ 
with a Wilson-loop holonomy, 
bions (instanton-monopoles) in Yang-Mills theory can exist inside the vortex 
as the ${\mathbb C}P^{N-1}$ bions (instanton-domain walls). 
By taking a un-Higgsing limit, the vortex disappears,  
and therefore we expect that they remain as  Yang-Mills bions.

The same relation holds between 
quark matter in high density QCD and 
the ${\mathbb C}P^{2}$ model on a non-Abelian vortex 
\cite{Balachandran:2005ev} 
(see Ref.~\cite{Eto:2013hoa} for a review). 
This may give a hint to understand a quark-hadron duality 
between the confining phase at low density and 
the Higgs phase at high density,  
through a non-Abelian vortex 
\cite{Eto:2011mk}.


\begin{acknowledgments}
T.\ M.\  is grateful to M.~\"{U}nsal and G.~Dunne for the fruitful 
discussion. T.\ M.\ appreciates the KMI special lecture ``Resurgence and 
trans-series in quantum theories" 
at Nagoya University and thanks the organizer T.~Kuroki. 
T.\ M.\ is thankful to T.~Iritani, E.~Itou, K.~Kashiwa and T.~Kanazawa for the discussion
on the related works.
T.\ M.\  is in part supported by the Japan Society for the Promotion of Science (JSPS)
Grants Number 26800147. 
The work of M.\ N.\ is supported in part by Grant-in-Aid for 
Scientific Research (No. 25400268) and by the ``Topological 
Quantum Phenomena''  Grant-in-Aid for Scientific Research on 
Innovative Areas (No. 25103720) from the Ministry of Education, 
Culture, Sports, Science and Technology  (MEXT) of Japan.
N.\ S.\  is supported by Grant-in Aid for Scientific Research 
No. 25400241 from the Ministry of Education, 
Culture, Sports, Science and Technology  (MEXT) of Japan. 

\end{acknowledgments}


\end{document}